\documentclass[preprint,Letter]{aastex}
\usepackage{epsf}

\def\kms{\hbox{\kern 0.20em km\kern 0.20em s$^{-1}$}}
\def\cm{\mbox{\kern 0.20em {\rm cm}\kern 0.20em}}
\def\K{\mbox{\kern 0.20em {\rm K}\kern 0.20em}}
\def\pc{\mbox{\kern 0.20em {\rm pc}\kern 0.20em}}

\def\msun{{\rm M}_\odot\ }

\newcommand {\apgt} {\ {\raise-.5ex\hbox{$\buildrel>\over\sim$}}\ }
\newcommand {\aplt} {\ {\raise-.5ex\hbox{$\buildrel<\over\sim$}}\ } 

\def\kB{k_{\rm B}}

\shortauthors{Koyama \& Ostriker}
\begin{document}
\title{Pressure Relations and Vertical Equilibrium in the Turbulent,
  Multiphase ISM}
\author{Hiroshi Koyama\altaffilmark{1,2} and Eve C. Ostriker\altaffilmark{1}}

\altaffiltext{1}{Department of Astronomy,
University of Maryland, College Park, MD 20742, USA;
hkoyama@astro.umd.edu, ostriker@astro.umd.edu} 
\altaffiltext{2}{Current address: High-Performance Computing Team,
Integrated Simulation of Living Matter Group, RIKEN,
61-1 Ono-cho, Tsurumi, Yokohama, 230-0046 Japan; hkoyama@riken.jp}

\begin{abstract}

We use numerical simulations of turbulent, multiphase,
self-gravitating gas orbiting in the disks of model galaxies to study
the relationships among pressure, the vertical distribution of gas,
and the relative proportions of dense and diffuse gas.  A common
assumption is that the interstellar medium (ISM) is in
vertical hydrostatic equilibrium.  We show that the disk height and
mean midplane pressure in our multiphase, turbulent simulations are indeed
consistent with effective hydrostatic equilibrium, provided that the
turbulent contribution to the vertical velocity dispersion and the
gas self-gravity are included. 
Although vertical hydrostatic equilibrium gives a
good estimate for the mean midplane pressure $\langle P\rangle_{\rm
midplane}$, this does not represent the pressure
experienced by most of the ISM.  Mass-weighted mean pressures $\langle
P \rangle_{\rho}$ are typically an order of magnitude higher than
$\langle P\rangle_{\rm midplane}$ because self-gravity concentrates
gas and increases the pressure in individual clouds without raising
the ambient pressure.

We also investigate the ratio $R_{\rm mol}=M_{\rm
H_2}/M_{\rm HI}$ for our hydrodynamic simulations.
\citet{2006ApJ...650..933B} showed that $R_{\rm mol}$ is proportional
to the estimated midplane pressure in a number of systems.  We find
that for model series in which the epicyclic frequency $\kappa$ and
gas surface density $\Sigma$ vary together as $\kappa \propto \Sigma$,
we recover the empirical relation.  For other model series in which
$\kappa$ and $\Sigma$ are varied independently, the midplane pressure
(or $\Sigma$) and $R_{\rm mol}$ are not well correlated.  We conclude
that the molecular fraction -- and hence the star formation rate -- of a
galactic disk inherently depends on its rotational state, not just the local
values of $\Sigma$ and the stellar density $\rho_{\ast}$.  The empirical
result $R_{\rm mol} \propto \langle P\rangle_{\rm midplane}$ implies
that the three ``environmental parameters'' $\kappa$, $\Sigma$, and
$\rho_{\ast}$ are interdependent in real galaxies, presumably as a
consequence of evolution: real galaxies trend toward states with 
Toomre $Q$ parameter near unity.
Finally,
we note that $R_{\rm mol}$ in static comparison models far exceeds
both the values in our turbulent hydrodynamic simulations and observed
values of $R_{\rm mol}$, when $\Sigma > 10 \msun \ \pc^{-2}$,
indicating that incorporation of turbulence is crucial to obtaining a
realistic molecular fraction in numerical models of the ISM.
\end{abstract}

\section{Introduction}

All phases of the interstellar medium (ISM) are turbulent, and this
turbulence has many effects.  In the astrophysical literature,
turbulence is often treated as yielding a simple addition to the
thermal pressure, $P_{\rm total}= \rho(c_s^2 + v^2_{\rm turb})$, where
$v^2_{\rm turb}$ is the dispersion in the (one-dimensional) turbulent
velocity, and $c_s^2=P/\rho= f\kB T/\mu$ for gas with a total number
density $f n$ and mass density $\mu n$.  This approach is often
adopted when analyzing the stratification of interstellar gas clouds
and the ISM as a whole, with the combined pressure gradients taken to
balance the gravitational force per unit volume such that hydrostatic
equilibrium is maintained by the total pressure.  The turbulent
pressure is believed to be especially important in the cold components
of the ISM, for which observed linewidths far exceed the values of
$c_s$ inferred from excitation of atomic and molecular lines.

Models of effective hydrostatic equilibrium in the vertical direction, usually
assuming the turbulent and thermal velocity
dispersions are constants independent of height, are often applied to
observations of the large-scale Galactic ISM, and to observations of
the ISM in external galaxies (e.g.
\citealt{1991ApJ...382..182L,1994ApJ...433..687M,1995ApJ...448..138M,
1997A&A...326..554C,2000MNRAS.311..361O,
2002A&A...394...89N,2004ApJ...608..189D,
2004ApJ...612L..29B,2006ApJ...650..933B, 2008AstL...34..152K}).  For
example, \cite{2002A&A...394...89N} showed that the observed atomic
and molecular disk thicknesses in the Milky Way can be fit well by
assuming effective hydrostatic equilibrium, and accounting for both the gas
self-gravity and the external gravitational potential of stars and
dark matter.  \cite{2004ApJ...612L..29B} and
\cite{2006ApJ...650..933B} (hereafter BR06) used a simplified approach
to hydrostatic equilibrium in order to estimate the midplane gas
pressure in a sample of disk galaxies, adopting a single velocity
dispersion for the gas, treating the gravitational potential as
dominated by the stars, and assuming the stellar disk's scale height
is independent of radius.  \cite{2008AstL...34..152K} extended the
analysis of BR06 but instead of adopting a constant scale height for
the stellar disk, they assumed that the velocity dispersion for the
stars is consistent with a state of marginal gravitational instability
(with Toomre parameter $Q_{\ast}=1.5$) for the corresponding stellar surface
density.  They then assumed hydrostatic equilibrium for all (gaseous
and stellar) components separately, and computed the self-consistent
midplane pressure, finding differences of order $30-40\%$ from the
simplified BR06 approach.  Although widely adopted, the effective
hydrostatic equilibrium model for the large-scale ISM has not, to our
knowledge, been explicitly verified using actual turbulent flows.  One
of the goals of this work is to test this formulation systematically,
using the solutions of time-dependent numerical hydrodynamic
simulations of turbulent, multiphase gas.  

In addition to providing support against gravity, pressure also
affects the phase balance in the ISM.  For a static system at a given
mean density $\bar n$, changing the pressure alters the proportions of
mass divided between dense clouds and diffuse intercloud medium;
e.g. for cold and warm components in pressure equilibrium, the mass
ratio of cold to warm gas is 
$M_{\rm cold}/M_{\rm warm}=[\bar n/n_{\rm warm} - 1]/[1 - \bar
  n/n_{\rm cold}]=
[\bar n k T_{\rm warm} - P]/[P - \bar n k T_{\rm cold}]$. The mean
density itself, however, depends on pressure through the
condition of vertical hydrostatic equilibrium. Turbulent pressure, as
it affects the response to external and self-gravity, can be expected
to change both the mean density and the mass fractions of dense and
diffuse gas.  Here, we investigate these effects quantitatively.

The fraction of ISM mass in dense gas is important from the point of
view of galactic evolution, since this component is the immediate
precursor to star formation.  A recent observational study of external
disk galaxies by BR06 identified a linear relationship between the
mean ratio of molecular-to-atomic mass, $R_{\rm mol}$, and an estimate
for the total midplane pressure $\propto \sqrt{\rho_{\ast}} \Sigma$, where
$\rho_{\ast}$ is the stellar volume density and $\Sigma$ is the total
gaseous surface density.  BR06 propose that the molecular fractions in
widely-varying types of galaxies -- and hence their respective star
formation efficiencies -- are therefore determined essentially by a
single parameter, the midplane pressure.  To investigate this
proposal, we use multiphase turbulence simulations in which we
independently vary the input galactic ``environmental'' parameters.
The observational study of BR06 focused on the dependence of $R_{\rm
mol}$ on $\rho_{\ast}$ and $\Sigma$, but another important -- and
independent -- environmental
parameter is the angular rotation rate $\Omega$ (and the associated
epicyclic frequency $\kappa^2= R^{-3}d (\Omega^2 R^4)/dR$).  Using our
data sets from turbulent simulations, we compare the pressure estimate
of BR06 to the true value of the pressure, and also test how $R_{\rm
mol}$ relates to the mean pressure measured in two different ways.

We note that a number of recent numerical studies have investigated
the formation of ISM structures with internal densities reaching those
similar to Giant Molecular Clouds (GMCs).  Some studies 
\citep[e.g.][]{2002ApJ...564L..97K,2005ApJ...633L.113H,2006ApJ...648.1052H,2007A&A...465..431H,2007A&A...465..445H,2008A&A...486L..43H,2007ApJ...657..870V}
have focused
on how this may occur as a consequence of the collision of large-scale
high-velocity flows that shock and cool,
becoming turbulent at the same time.  Other studies have focused on
the ability of self-gravitating instabilities to induce converging
flows over sufficiently large scales that massive, high-column density
structures similar to observed GMCs are created
\citep[e.g.][]{2001ApJ...559...70K,2007ApJ...660.1232K,
2005ApJ...620L..19L,2006ApJ...639..879L}; these models include the
galactic shear and rotation that are important on these large scales,
and in some cases also include magnetic effects
\citep[e.g.][]{2002ApJ...581.1080K,2003ApJ...599.1157K}.  As spiral
arms are observed to be strongly associated with high molecular
fractions and star formation, some studies have focused on the
interaction between large-scale spiral shocks and self-gravity in
inducing GMC formation
\citep[e.g.][]{2002ApJ...570..132K,2006ApJ...646..213K,2008MNRAS.tmp.1249D}.
The details of conversion from diffuse to dense gas by cooling
downstream from spiral shock fronts has also recently been studied in
the absence of self-gravity
\citep[e.g.][]{2008MNRAS.389.1097D,2008ApJ...681.1148K}.  Taken
together, these and other recent studies have shown that significant
quantities of dense gas form naturally as a result of large-scale ISM
dynamical processes.  Of course, dense gas in the ISM 
is also returned to the
diffuse phases by the energetic inputs from star formation.  In the
present work, by incorporating feedback, we are able to evolve our
models until a quasi-steady state is reached.  This enables an
analysis of the correlations among statistical properties of the
system, in terms of their influence on the fraction of dense gas when
the system has reach a quasi-steady state of cloud formation and
destruction.

This paper is organized as follows: In \S 2 we briefly summarize our
numerical methods.  The specification of model parameters and the
results of statistical analysis in comparison to the
vertical-equilibrium approximation are presented in \S 3.  In \S 4, we
discuss the molecular fraction and investigate how it relates to the
ISM pressure in our models.  We summerize our results and discuss
implications for ISM structure and evolution in \S 5.

\section{Numerical Methods}

The analysis in this paper are based on time-dependent numerical
hydrodynamic simulations of turbulent, multiphase, interstellar gas.
Details of our numerical methods are presented in a companion paper
(Koyama \& Ostriker 2008, hereafter Paper I); here, we
briefly summarize the model properties and parameterizations.  The
models we use are two-dimensional, representing slices through the
ISM in radial-vertical ($R-z$) planes.  We include sheared galactic
rotation, a radial gravitational force (the centrifugal force and
gravity balance in the
unperturbed state, which assumes a rotation curve $V_c=const$), and
Coriolis forces in the equations of motion, as well as gaseous
self-gravity and vertical gravity representing the potential of the
stellar disk.  The gas is treated as a single fluid in chemical
equilibrium, and we include (volumetric) radiative heating and cooling
processes as a function of density and temperature appropriate to the
range $10 < T < 10^4$ K.  The thermal processes we incorporate include
photoelectric heating from small grains and polycyclic aromatic
hydrocarbons, heating and ionization by cosmic rays and X-rays,
heating by H$_2$ formation and destruction, atomic line cooling from
Hydrogen Lyman $\alpha$, CII, OI, Fe II, and Si II, rovibrational line
cooling from H$_2$ and CO, and atomic and molecular collisions with
grains.  We adopt shearing-periodic boundary conditions in the radial
direction.

To drive turbulence, we also include a model of stellar feedback:
within ``HII regions'' (which are defined by contours of the perturbed
gravitational potential surrounding regions where the density has
exceeded a specified threshold), the gas heating rate is increased by
a factor 1,000. As a consequence, gas within these ``HII regions''
heats to temperatures $\sim 10^4$K, irrespective of density.  The
detailed recipe for the feedback phenomenon is described in
\citeauthor{Paper1}. Our aim is not to represent star formation
feedback in a fully realistic manner, but to drive turbulence in a way
similar to that which occurs within the dense
ISM.  In this sense, our feedback prescription is similar
in spirit to simulations of giant molecular clouds in which turbulence
is applied via arbitrary forcing functions (e.g.
\citealt{1998ApJ...508L..99S,1999ApJ...524..169M,
2000ApJ...535..887K}).  Thus, our results should be taken as
demonstrating the physical importance of turbulence to setting
properties such as the vertical thickness of the disk, not as giving
quantitative predictions for what the value of the disk thickness,
etc., should be.

\section{Model Series and Results}

In our local disk models, three free parameters are needed to
characterize the ``galactic environment'': the total surface
density of the gas $\Sigma$, the local epicyclic frequency $\kappa$, 
and the local stellar density $\rho_{\ast}$.  
As we assume a flat rotation curve, $\kappa=\sqrt{2}\Omega$ where $\Omega$ is
the angular rotation rate at the center of our domain.  The stellar
density is used in order to specify the vertical gravity 
${\bf g}_{\ast} =- 4 \pi G \rho_{\ast} z \hat z$.  

Following \citeauthor{Paper1}, we study four Series of models to explore
the parameter dependence of our results. 
For each Series, we hold two quantities fixed and vary a third
quantity, as follows:
\begin{itemize}
\item Series Q: $\kappa/\Sigma$ and
      $\sqrt{\rho_{\ast}}/\Sigma$ are constant while $\Sigma$ varies;
\item Series K: $\kappa$ and $\sqrt{\rho_{\ast}}/\Sigma$ are constant
  while $\Sigma$ varies;
\item Series R: $\kappa/\Sigma$ and $\rho_{\ast}$ are constant while 
$\Sigma$ varies;
\item Series S: $\Sigma$ and $\rho_{\ast}$ are constant while $\kappa$
  (and $\Omega$) varies.
\end{itemize}
Since Toomre's parameter is proportional to $\kappa/\Sigma$,
Series Q and R would have constant gaseous $Q=\kappa c_s/(\pi G \Sigma)$ if
the sound speed $c_s$ were constant.  
The Q and R series correspond to values of $Q=2.1 (c_s/7\kms)$. 
Assuming a constant stellar velocity
dispersion, $\Sigma_{\ast} \propto \sqrt{\rho_{\ast}}$, so that 
the stellar Toomre parameter 
(hereafter $Q_{\ast}$) would also have the same value for all members of 
Series Q.  
In all members of the R and S Series and in the 
$\Sigma=15.0 ~\msun\pc^{-2}$ models
of the Q and K series, we take 
$\rho_{\ast}=0.14 ~\msun\pc^{-3}$.  
In the K Series, we use 
$\kappa=62.4 ~{\rm km}~{\rm s}^{-1}{\rm kpc}^{-1}$, 
 while in the S Series we use $\Sigma=15.0 ~\msun\pc^{-2}$.

This paper focuses on how turbulence affects the vertical structure of
the galactic ISM.  An important aspect of our studies is to understand
how the results differ from the situation in which turbulence is
absent.  Thus, as baselines for comparison, we have two vertical
non-turbulent model Series: one in which the gas and stellar surface
densities are proportional (Series HSP), and one in which the stellar
surface density is constant (Series HSC).  These correspond to
dynamical Series Q and K (for HSP) and Series R (for HSC),
respectively.  These models are one-dimensional in the vertical ($z$)
direction; each model represents the asymptotic hydrostatic
equilibrium state which develops in the absence of any stellar
feedback.

\begin{figure}
\vskip 2in

\epsscale{1.5}
\plottwo{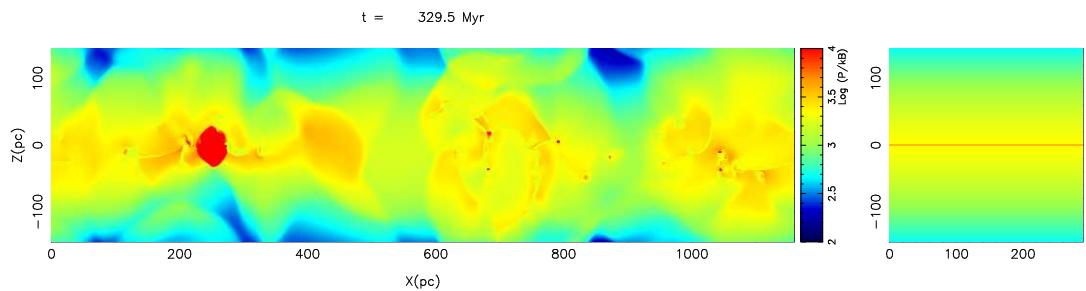}{p2f1bcolor.ps}
 \caption{ {\it Left}: A snapshot of gas pressure (logarithmic color
 scale) from Model Q11 simulation.  For comparison, the {\it right} panel 
shows the pressure in the hydrostatic model (HSP11) that has the same
 total gas surface density $\Sigma$ and stellar density $\rho_*$ 
as Model Q11. }
\label{fig:snapshot}

\vskip 4in
\end{figure}

Figure \ref{fig:snapshot} shows a snapshot of the gas 
pressure in a dynamical model from Series Q, compared to the
hydrostatic model from Series HSP.  The density and temperature are
shown for the same snapshot in Figure 1 of Paper I. In the dynamical
model, the pressure overall increases toward the midplane, but there
are significant variations associated with structure in the gas; for
the particular snapshot shown, there is also a high-pressure region
near the left of the figure, which is associated with a locally-heated
star formation feedback region.  The hydrostatic model shows a secular
increase in pressure towards the midplane.

\subsection{Vertical Scale Height
\label{scale_height}
}

\begin{figure}
\epsscale{1.0}
\plottwo{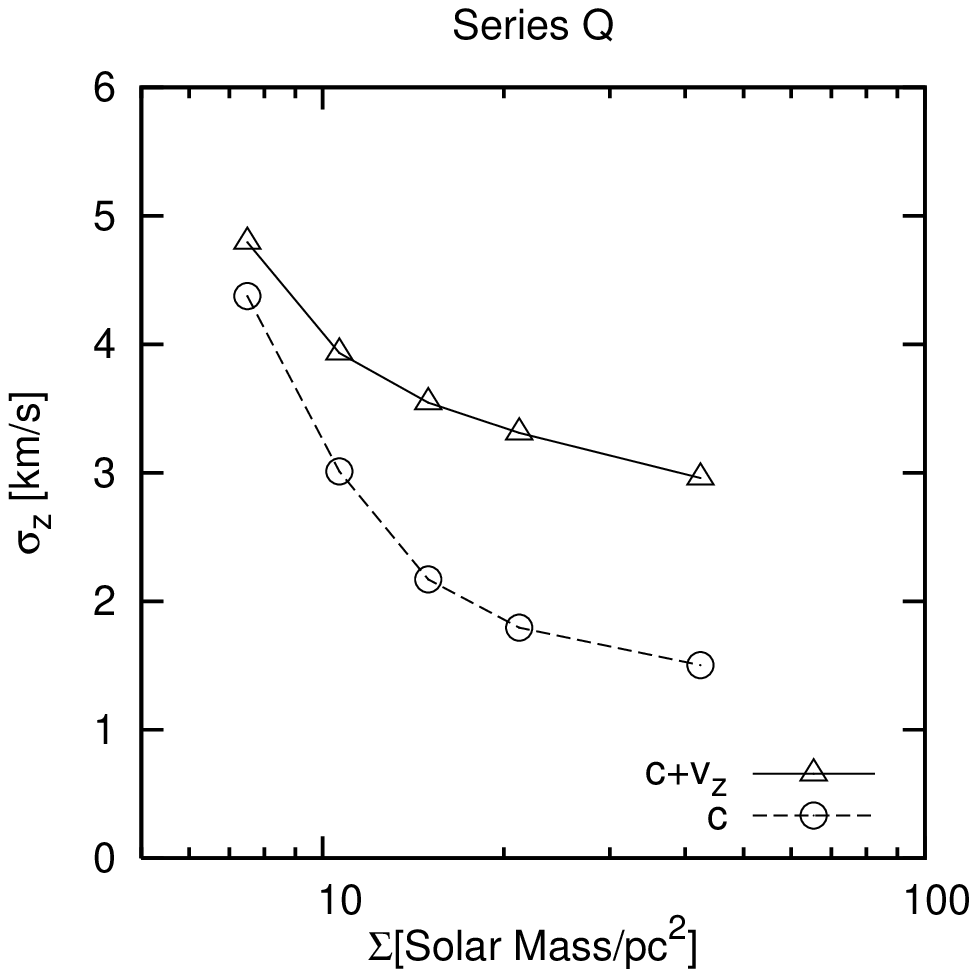}{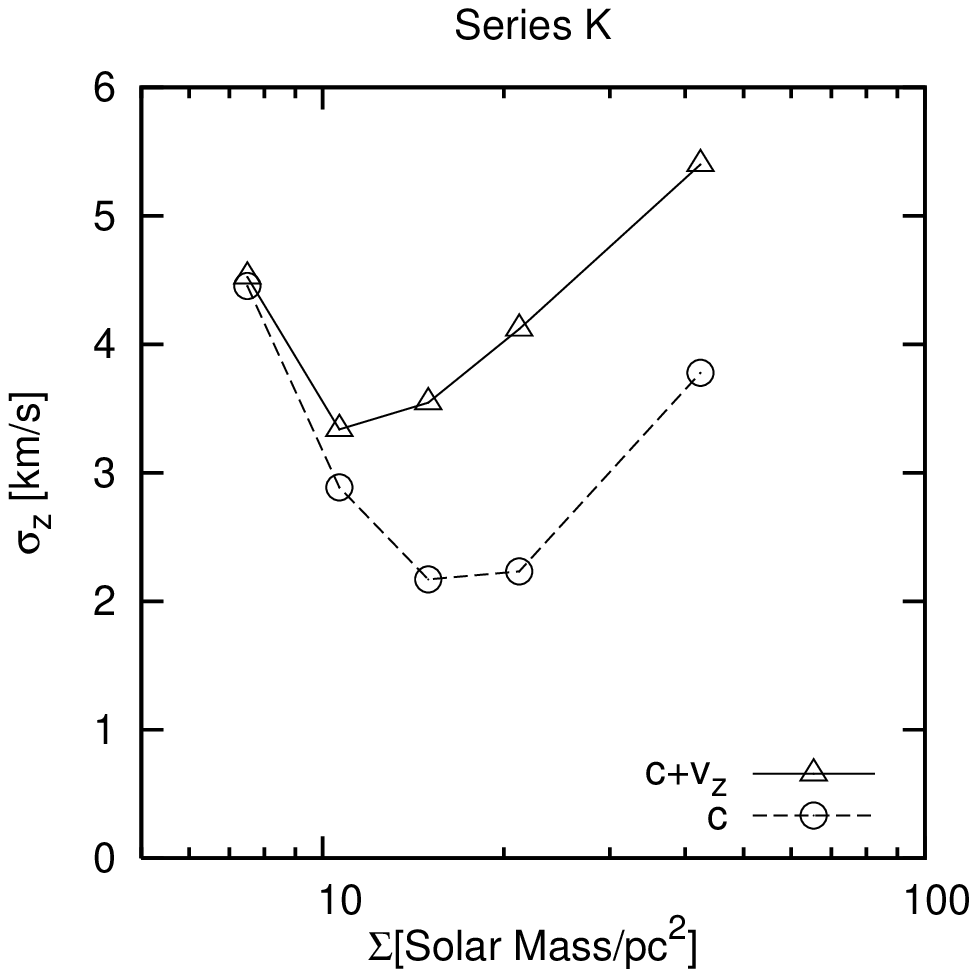}

\plottwo{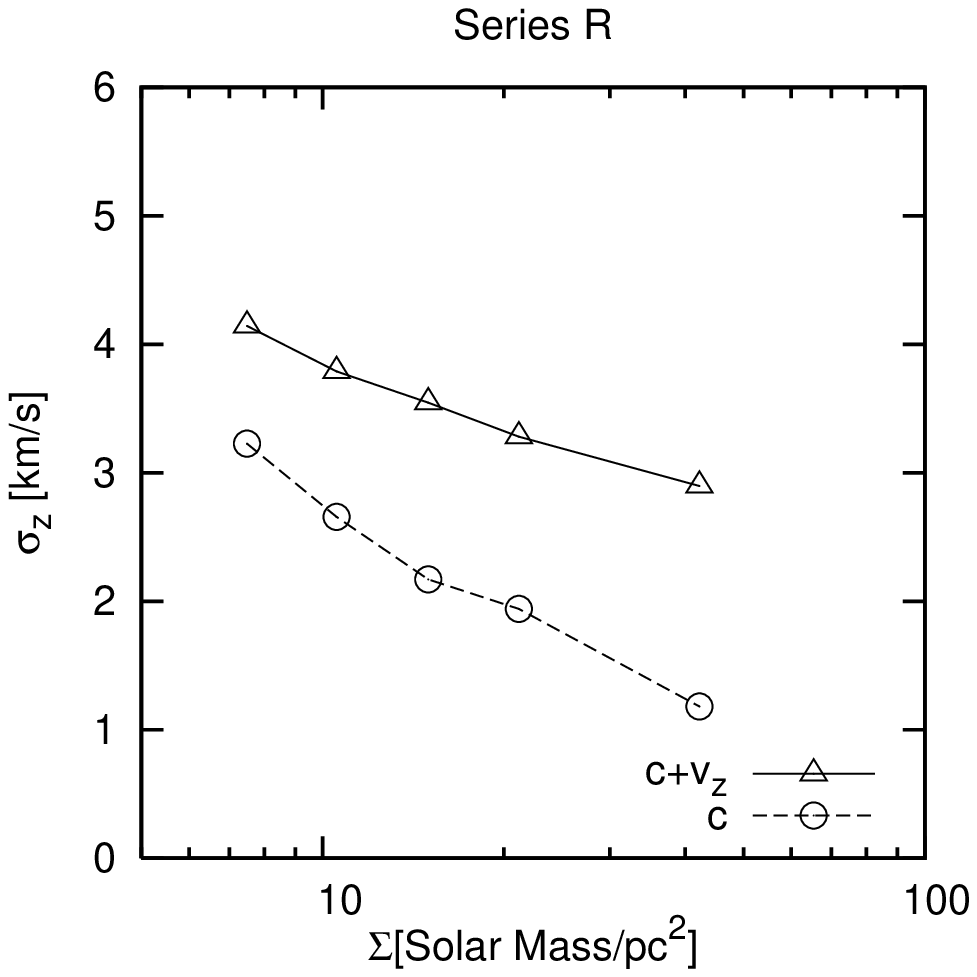}{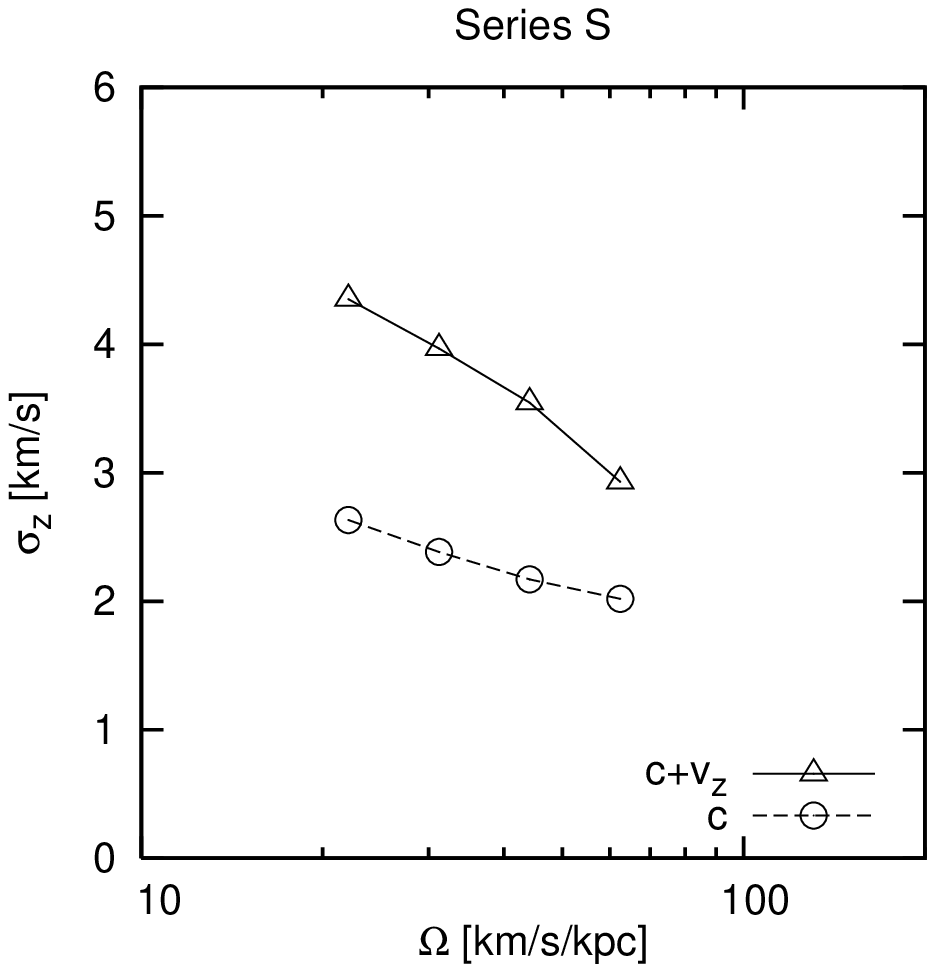}
 \caption{Mean vertical velocity dispersion, weighted by mass.  Both
the thermal $c_s$ ({\it circles}) and the total (thermal + turbulent)
$\sigma_z=\sqrt{c_s^2+v_z^2}$ ({\it triangles}) dispersions are shown
for all Series.}
\label{fig:vz}
\end{figure}

We begin by examining the vertical velocity dispersion of gas in all
of the model Series.  Figure \ref{fig:vz} shows space- and
time-averages (weighted by mass) of both the thermal velocity
dispersion $c_s=(P/\rho)^{1/2}$ ({\it circles}) and the combined
thermal + turbulent velocity dispersion $\sigma_z=\sqrt{c_s^2+v_z^2}$
({\it triangles}).  The four panels correspond to the Series Q, K, R,
and S.  In Series Q and R, the mean thermal velocity dispersion
decreases with increasing surface density. The reason for this is that
the mass fraction of cold, dense gas increases with $\Sigma$ in all of
these models (see \citeauthor{Paper1}). This is because gravity is
lower, and gas is less compressed (both vertically, and horizontally
by self-gravity), at low $\Sigma$.  In Series K, on the other hand,
the mean thermal speed has a local minimum at intermediate $\Sigma$.
Again, this can be understood in terms of the mass fraction of warm
gas, which is largest at low and high $\Sigma$ (see
\citeauthor{Paper1}) in this Series; at high $\Sigma$, the model is
extremely active in terms of feedback because (with constant $\kappa$)
the disk is quite unstable gravitationally.  For all the series in
which $\Sigma$ is the variable parameter (i.e. Q, K, and R), the
turbulent part of the total velocity dispersion increases with
$\Sigma$; this is because the higher-$\Sigma$ models have higher
feedback rates, and therefore increasing (or flat) turbulence levels.
For Series S (with constant $\Sigma$), the turbulence decrease as
$\Omega$ increases, as high $\kappa$ stabilizes the disk and prevents
gravitational collapse and feedback (see Figure 11 in
\citeauthor{Paper1}).  For all series, the (mass-weighted) turbulent
vertical velocity dispersion approaches or exceeds the (mass-weighted)
thermal velocity dispersion for some part of parameter space, so that
turbulent support of gas in the vertical gravitational field is
expected to be important.

\begin{figure}
\plottwo{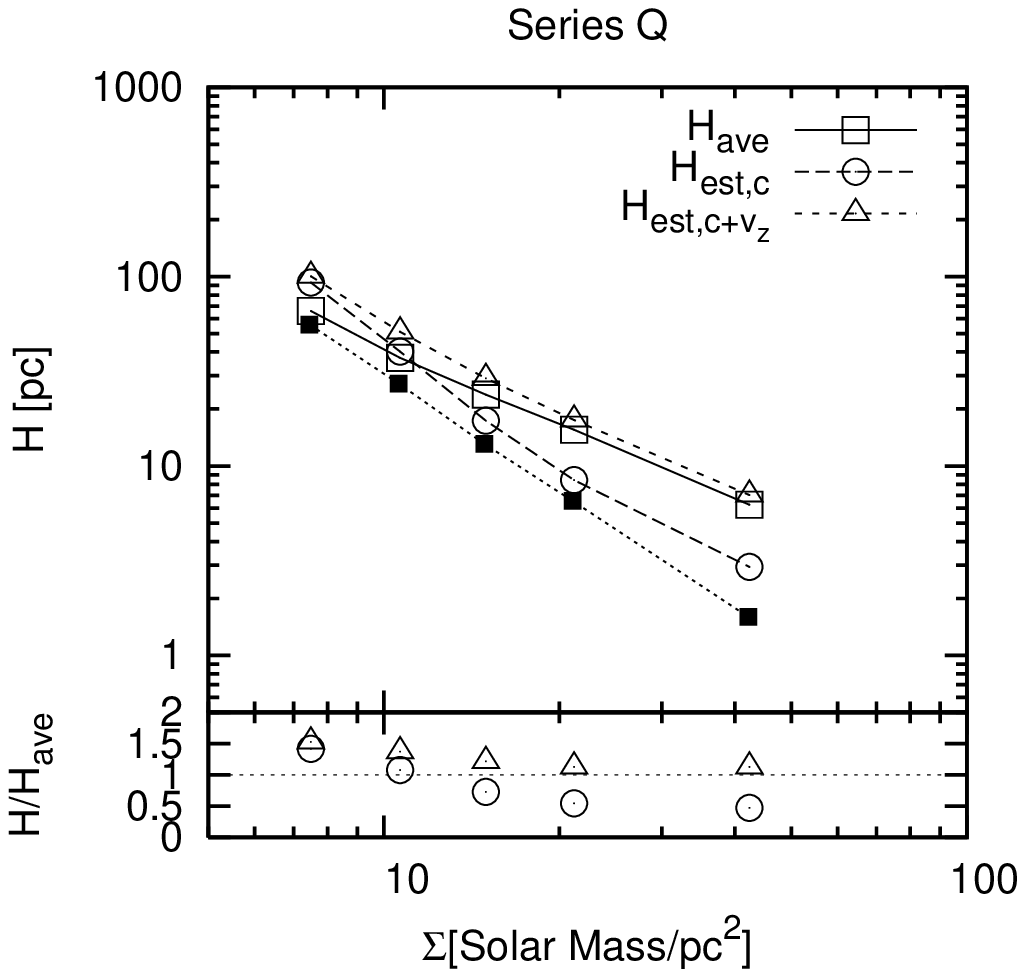}{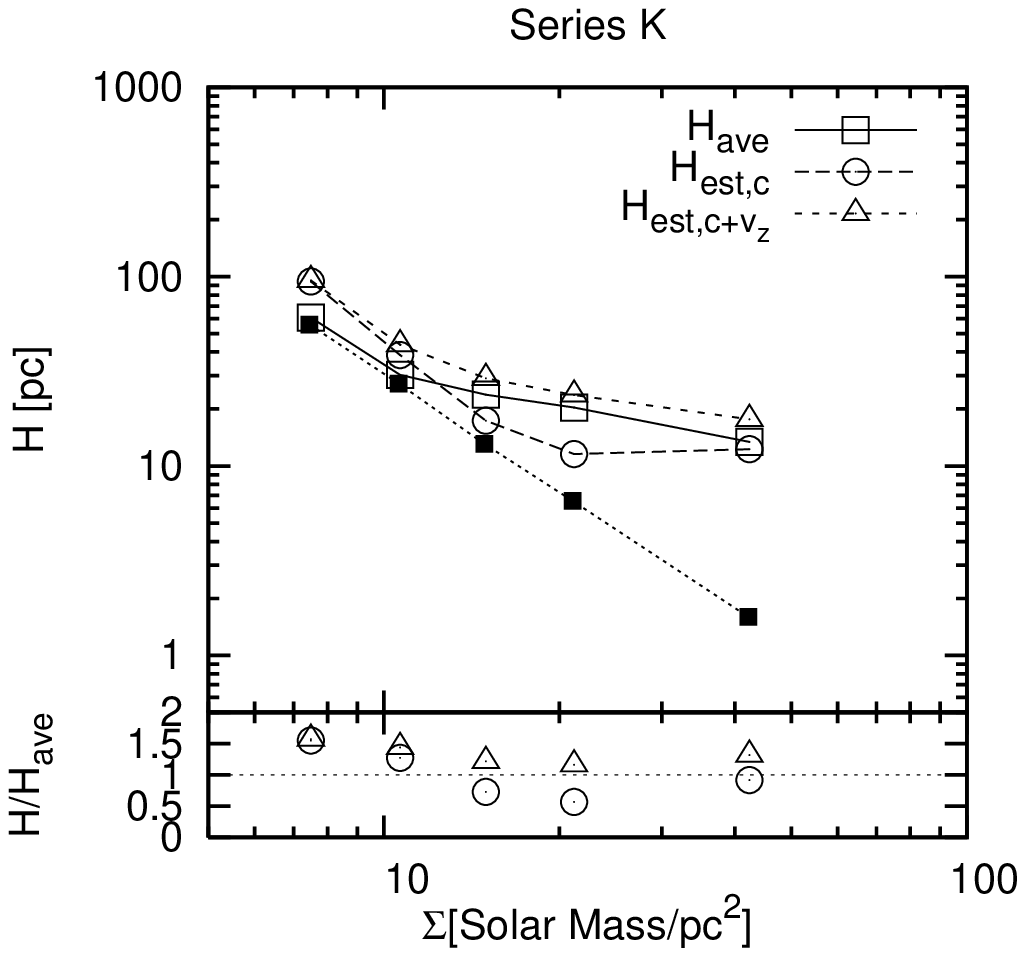}

\plottwo{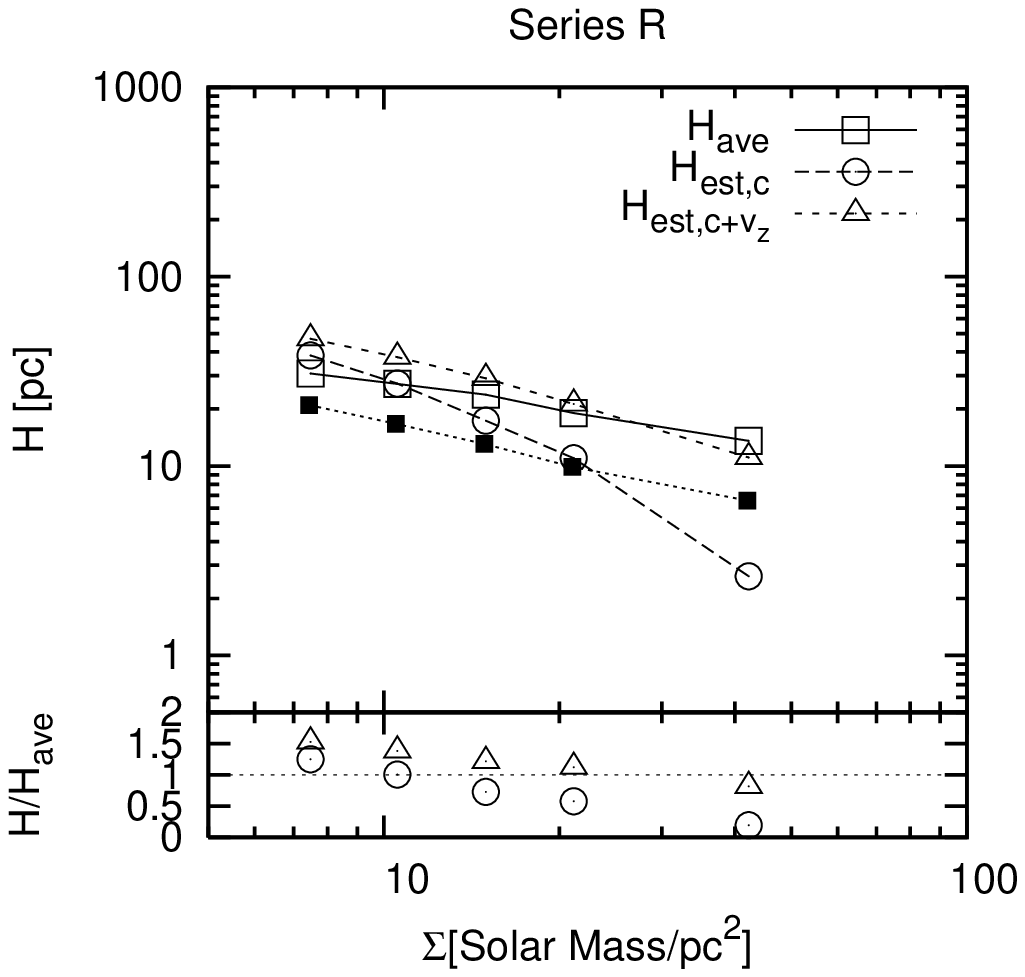}{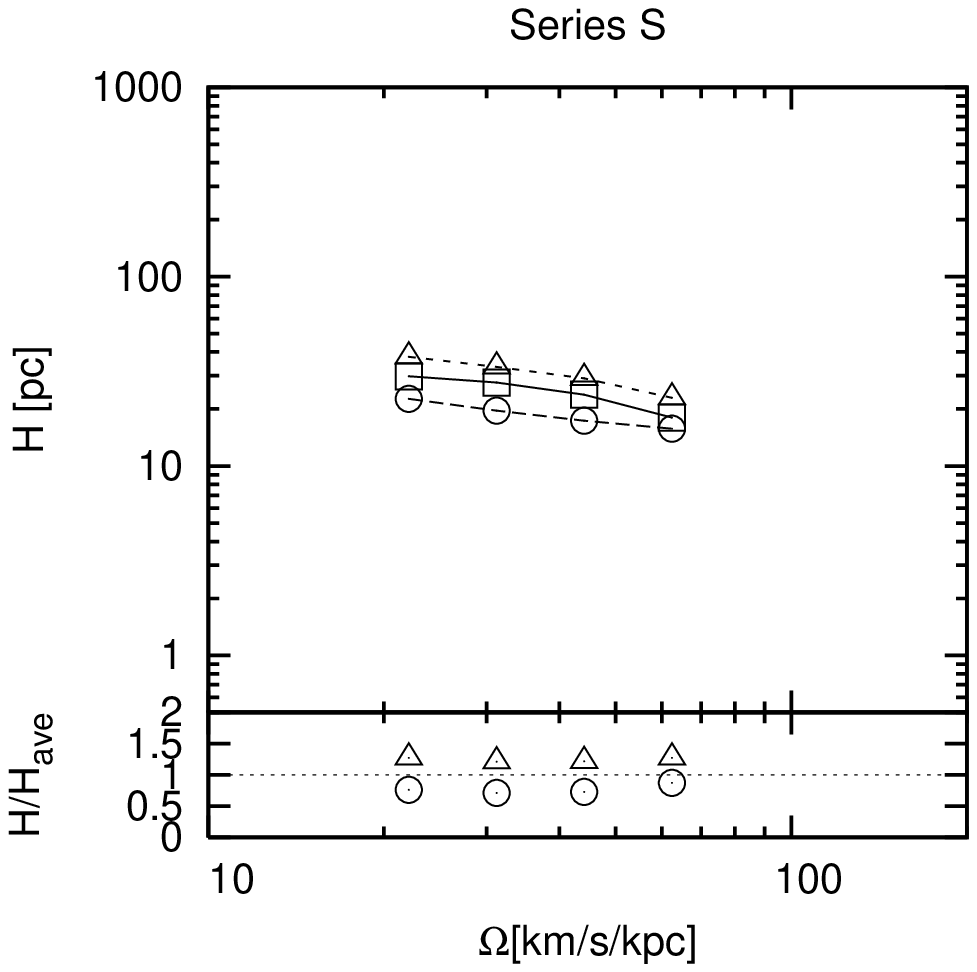}
 \caption{Disk scale heights, for all hydrodynamic and hydrostatic models.
{\it Open boxes} denote the directly-measured scale height (see
eq. \ref{Have_eq}) for all hydrodynamic Series.
{\it Filled boxes} show the measured scale height for corresponding 
hydrostatic models (HSP for Series K and Q, HSC for Series R). 
{\it Open circles} and {\it triangles} 
show the estimated scale heights (see eq. \ref{Hest_eq}) using thermal and
 thermal plus turbulent velocity for $\sigma_z$, respectively. 
The bottom part of each panel shows the ratio of estimated scale heights to
direct measurements. 
}\label{fig:Have}
\end{figure}

Next, we measure (for all Series) the vertical scale height, using 
the following averaging:
\begin{eqnarray}
H_{\rm ave}=\sqrt{\frac{\sum_{\rm all ~ zones} \rho z^2 }
{\sum_{\rm all ~ zones} \rho}}
\label{Have_eq}
\end{eqnarray}
where $z$ is the vertical coordinate relative to the midplane.  We
further average the values of $H_{\rm ave}$ over time.
In order to test whether the velocity dispersion can be used to
obtain an accurate measure of the scale height, we also compute 
``estimated'' vertical scale heights defined as:
\begin{eqnarray}
H_{\rm est}&=&
\frac{1}{\sqrt{2\pi}}
\frac{\sigma_z^2}
{G\Sigma+\left[(G\Sigma)^2+2G\rho_{\ast}\sigma_z^2\right]^{1/2}}
=
\frac{\sigma_z}{\sqrt{4\pi G \rho_*}}
\frac{1}
{A+\left[A^2+1\right]^{1/2}}\label{Hest_eq} \\
&=&\left\{\begin{array}{ll}
\displaystyle \frac{\sigma_z}{\sqrt{4\pi G\rho_{\ast}}} & (\Sigma \to 0)\\
\displaystyle \frac{\sigma_z^2}{\sqrt{8\pi}G\Sigma} & (\rho_{\ast} \to 0).
\end{array}
\right.
\end{eqnarray}
This formula (see Appendix for derivation) accounts for both gas
self-gravity and stellar gravity; the limiting forms are for
negligible gaseous and stellar gravity, respectively.  

In equation (\ref{Hest_eq}), 
$A$ is a dimensionless factor that measures the relative
densities of the gaseous and stellar disks,
\begin{eqnarray}
A\equiv\sqrt{\frac{G\Sigma^2}{2\rho_{\ast}\sigma_z^2}}
=\frac{\Sigma \, c_{\ast,z}}{\Sigma_{\ast} \, \sigma_z \sqrt{\pi}}.
\end{eqnarray}
The latter expression treats the stellar disk as an isothermal
self-gravitating equilibrium, with 
$H_{\ast}=c_{\ast,z}^2/(\pi G \Sigma_{\ast})$,
and shows that $A\sim Q_{\ast}/Q$ (assuming that vertical and radial
velocity dispersions are proportional).  
The formula (\ref{Hest_eq}) may be thought of as an extension of
the usual non-self-gravitating scale height formula to account for the
gravity of the gas.  Since $A>0$, the correction factor depending on
$A$ is always $<1$.  
If the gas disk is much more gravitationally unstable than
the stellar disk ($A \sim Q_{\ast}/Q \gg 1$), the correction factor is
large; otherwise the correction factor is order-unity.

Figure \ref{fig:Have} shows the measured ($H_{\rm ave}$) and ``predicted''
($H_{\rm est}$) disk scale heights for all series of hydrodynamic
models.  For $H_{\rm est}$, we show results using for $\sigma_z$
either the thermal velocity dispersion ($\sigma_z=c_s$; subscript $c$)
or the total velocity dispersion ($\sigma_z^2=c_s^2+v_z^2$; subscript
$c+v_z$).  To show how turbulence contributes to setting the disk
thickness, $H_{\rm ave}$ is also shown for the hydrostatic models.
The difference between $H_{\rm ave}$ in hydrostatic and hydrodynamic
models can be quite large, up to a factor 10 in some cases.  We note
that $H_{\rm ave}$ of the hydrostatic models ({\it filled
boxes}) differs from $H_{{\rm est},c}$ ({\it open circles}) because the
mass-weighted mean sound speed differs for hydrostatic and
hydrodynamic models.

Overall, Figure \ref{fig:Have} shows that the estimate for scale
height $H_{{\rm est},c+v_z}$ that includes turbulence traces the
measured $H_{\rm ave}$ quite well, for all the Series.  The difference
between $H_{{\rm est},c}$ and $H_{{\rm est},c+v_z}$ increases with
increasing $\Sigma$, with quite large differences for some of the
models in Series Q and R.  This indicates that high surface density
disks are supported largely by turbulent velocities, in these cases.
To facilitate comparisons between estimated and measured value of the scale
height, in the lower part of each panel we also show the ratios
$H_{{\rm est},c }/H_{\rm ave}$ ({\it circles}) and $H_{{\rm est},c+v_z
}/H_{\rm ave}$ ({\it triangles}).  At low values of $\Sigma$ in Series
Q, K, and R, both estimates of $H$ exceed the true measured value.  It
is notable that where the turbulent contributions are large, at high
$\Sigma$ in Series Q, K, and R, the estimated and measured disk
thicknesses are in quite good agreement (within $\sim 10-20\%$).
Thus, we conclude that if measurements of the vertical
velocity dispersion together with the gaseous surface density and
stellar surface density can be made observationally, they can be
combined to yield an accurate estimate of the gas disk's thickness.

\subsection{Gas Pressure}

The gaseous pressure, like the scale height, is often difficult to
measure directly.  As a consequence, other proxies are often used to
obtain an estimate of the value of the pressure, with an assumption
that vertical equilibrium is satisfied.  Here, we
test how well such pressure estimates agree with the directly-measured
pressure, for our multiphase turbulent models.

\begin{figure}
\plottwo{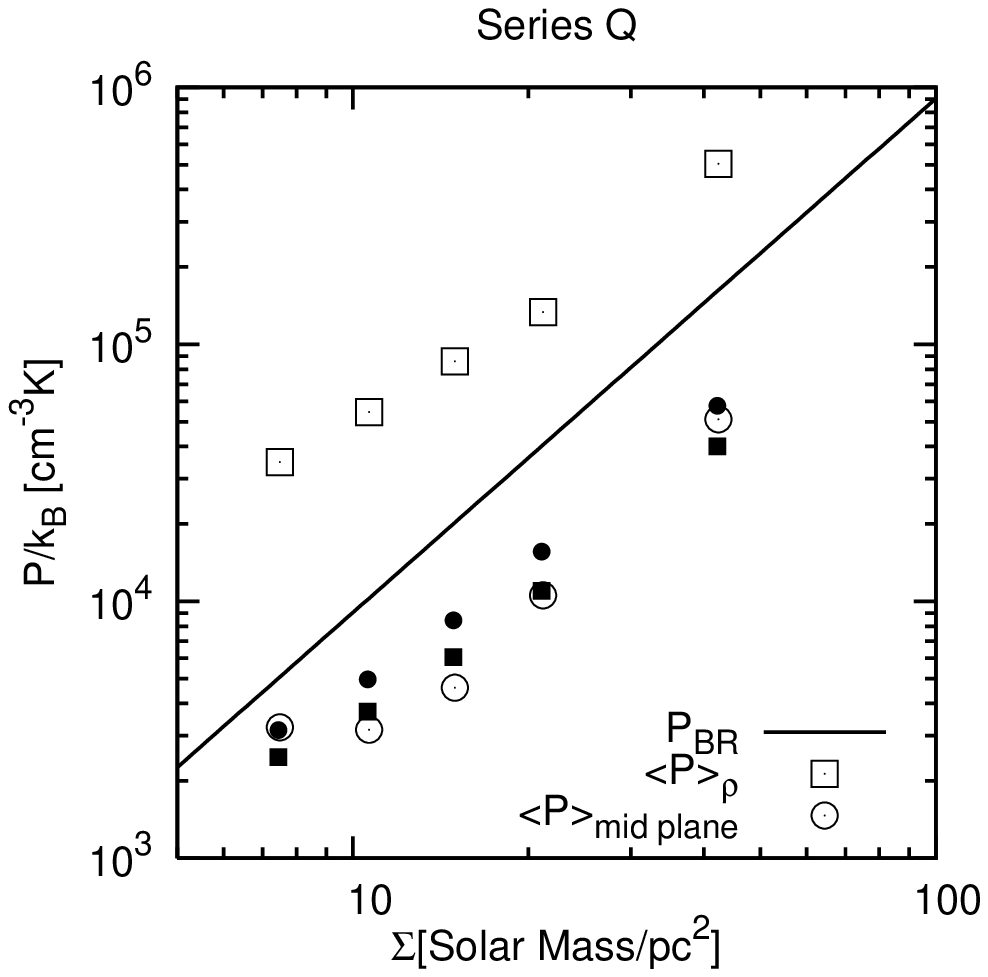}{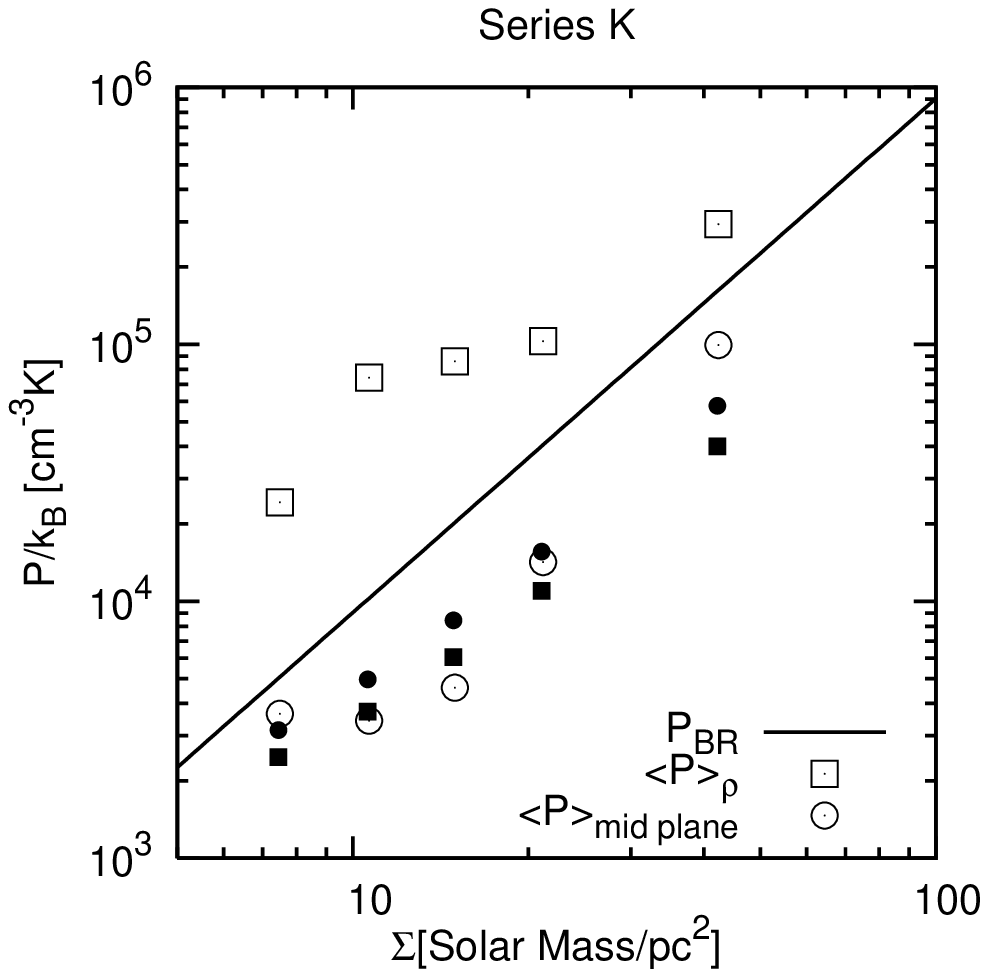}

\plottwo{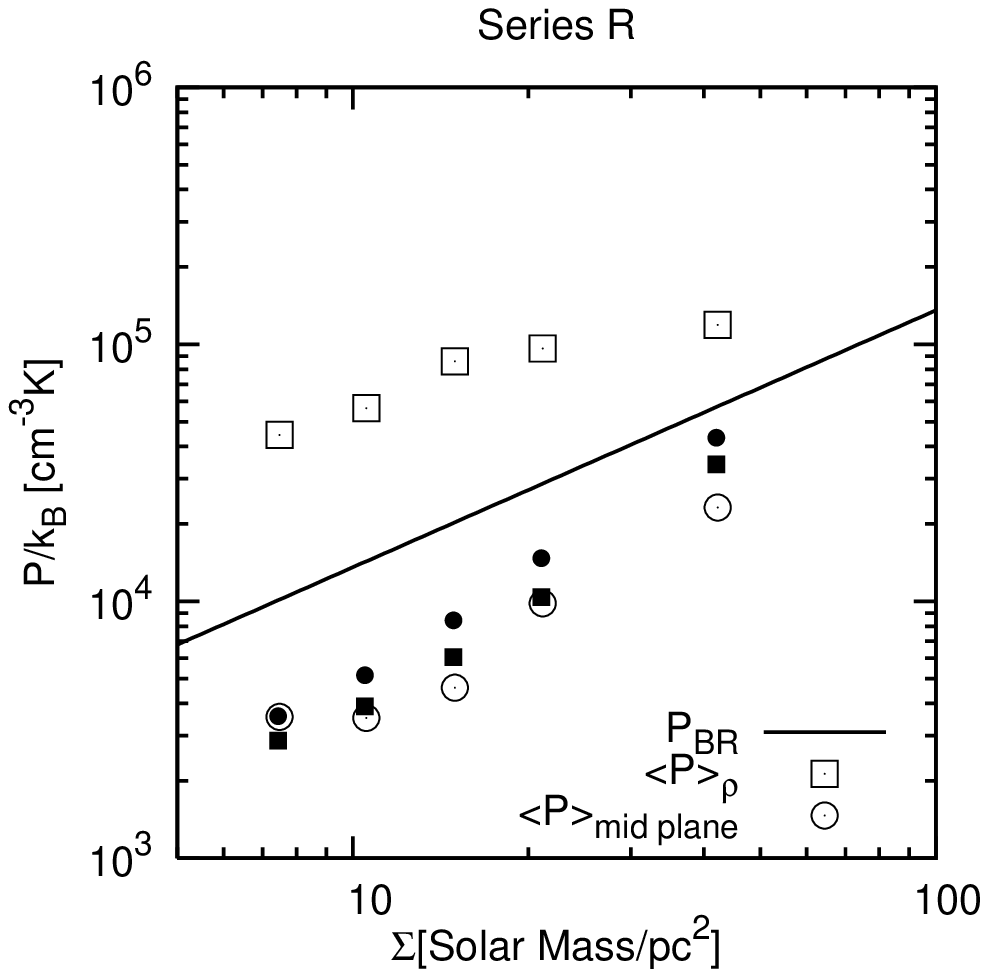}{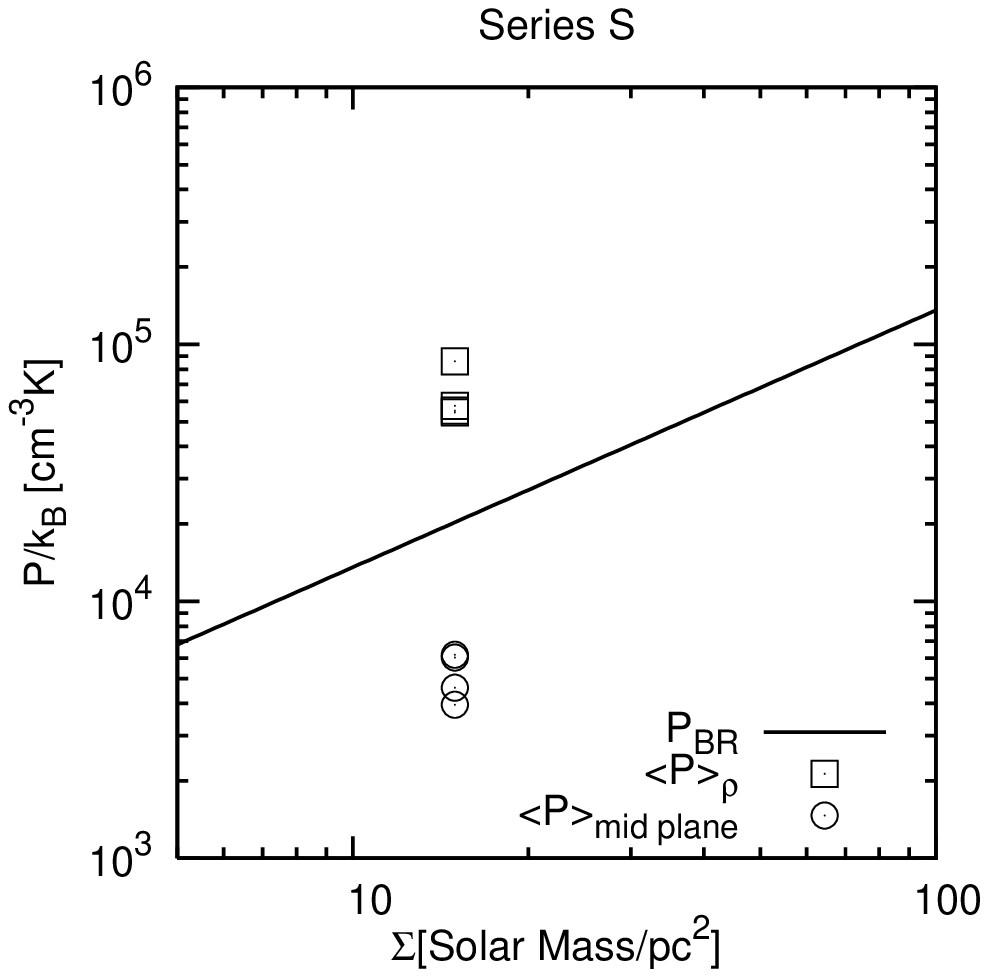}
 \caption{Gas pressure averages for all Series.
{\it Open boxes} show mass-weighted averages and {\it open circles}
show the midplane pressure, for hydrodynamic models. 
{\it Filled boxes} and {\it filled circles} show 
the same for hydrostatic models.
The pressure estimate of BR06 is also indicated {\it solid line}.}
\label{fig:Pressure}
\end{figure}

Figure \ref{fig:Pressure} shows for all models in all hydrodynamic 
Series the average gas pressure.
We consider two different ways of averaging:  
weighting by mass 
$\langle P \rangle_{\rho}$ ({\it open box}), and weighting by volume
$\langle P\rangle_{\rm midplane}$ ({\it open circle}).  
The value $\langle P \rangle_{\rho}$ is interesting because it
characterizes the value of pressure experienced by the average atom or
molecule, whereas $\langle P\rangle_{\rm midplane}$ is interesting 
because it represents the pressure in the diffuse
(non-self-gravitating) part of the ISM that is closest to star-forming
regions. 

The mass- and volume- weighted averages
are defined by the following: 
\begin{eqnarray}
\langle P \rangle_{\rho}&=&\frac{\int P dm}{\int dm},\\
\langle P \rangle_{\rm midplane}&=&\int 
\frac{P_{\frac{N_z}{2}}+P_{\frac{N_z}{2}+1}}{2} \frac{dx}{L_x}.
\end{eqnarray}
For $\langle P \rangle_{\rho}$, all zones in the domain are included,
while for $\langle P\rangle_{\rm midplane}$, the  
subscripts $\frac{N_z}{2}$ and $\frac{N_z}{2}+1$ indicate
that only zones in the two horizontal planes closest to the midplane
are included.
Time averaging is applied in all models after the above space averaging.
We also show the same pressure averages for the 
hydrostatic Series ({\it filled box} and {\it filled circle}).
Interestingly, in the hydrodynamic models 
$\langle P \rangle_{\rho}$ always exceeds 
$\langle P\rangle_{\rm midplane}$ by a large factor $\sim 10$.
This indicates that self-gravity is important in increasing the
pressure above the ``ambient'' value, for much of the gas.
Pressures cannot exceed the ambient midplane value without horizontal
gradients, which are balanced by the gravity within individual clouds 
(see Fig. \ref{fig:snapshot}).  
In the hydrostatic models, $\langle P \rangle_{\rho}$
({\it filled boxes})
is generally slightly below $\langle P\rangle_{\rm midplane}$
({\it filled circles}), because the pressure at the midplane is the
maximum within any system, and weighting by mass includes 
lower-pressure gas which reduces the average.  
(Note that for the hydrostatic models, there are no horizontal
gradients in any quantities; see Figure \ref{fig:snapshot}.)
Except in the most
active disks, the mass-weighted
averages for the hydrostatic models are close to the midplane values
for the hydrodynamic models.
In Figure \ref{fig:Pressure} we also
display the pressure estimate of BR06 ({\it solid line}) defined as: 
\begin{equation}
P_{\rm BR}=\Sigma v \sqrt{2G\rho_{\ast}},
\label{PBR_eq}
\end{equation}
where $v=8$ km/s is adopted. 
This line falls between $\langle P \rangle_{\rho}$ and 
$\langle P\rangle_{\rm midplane}$
for all the hydrodynamic Series. 

For hydrostatic Series HSP (shown in the Series Q and K panels), the
slope of the midplane pressure is close to that predicted by equation
(\ref{PBR_eq}), while being offset to lower $P$ by a factor 2-3.  The
difference in slope is because the medium has multiple phases, rather
than a single phase at a given thermal sound speed.  The offset is
because (i) much of the mass in the hydrostatic models is at low
temperatures, for which the sound speed is well below $8$ km/s, and
(ii) equation (\ref{PBR_eq}) does not include the gaseous vertical
gravity, which is comparable to the stellar gravity when vertical
velocity dispersion is low and the disk is very thin (see below).
These effects push $P$ in opposite directions, and hence partially 
compensate each other.  For
hydrostatic series HSC (shown in the Series R panel), the prediction
of equation (\ref{PBR_eq}) departs significantly from the slope of the
midplane pressure results, because in the HSC series (which has
$\rho_*$ constant) vertical gravity is strongly 
dominated by gas rather than the stellar component at large $\Sigma$.

\begin{figure}
\plottwo{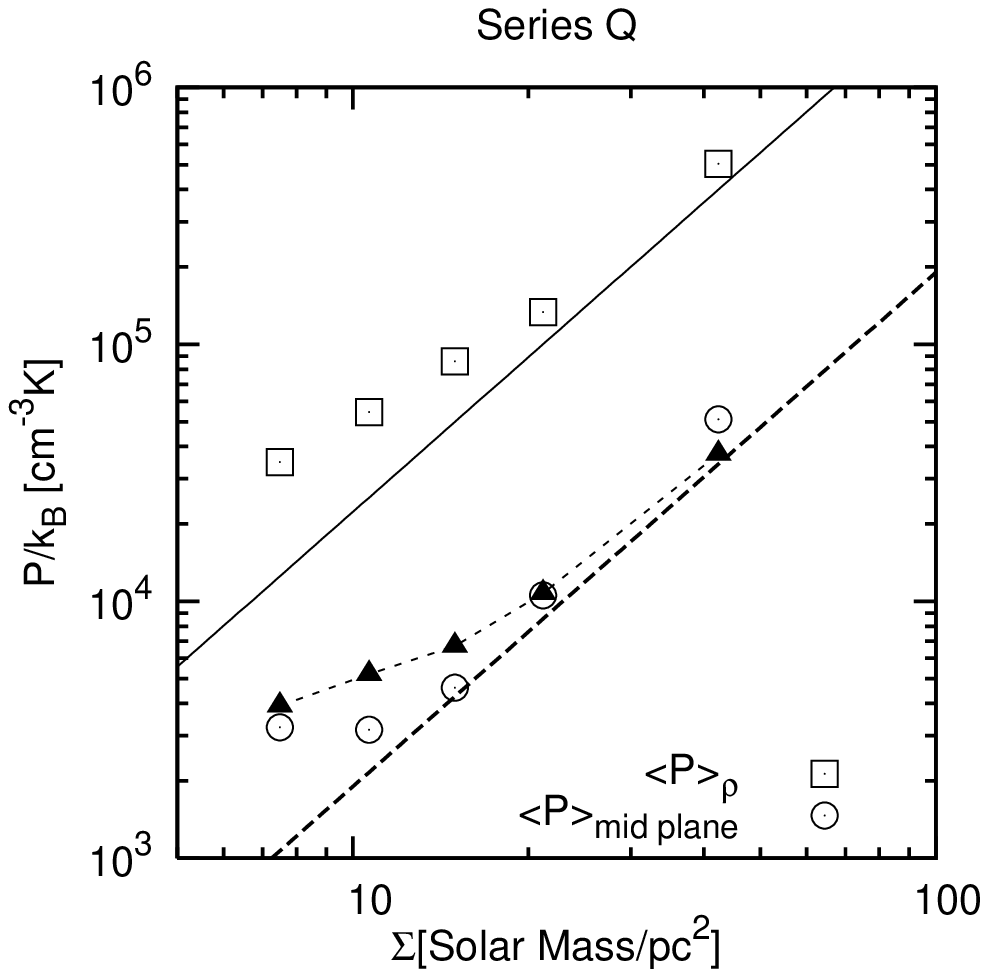}{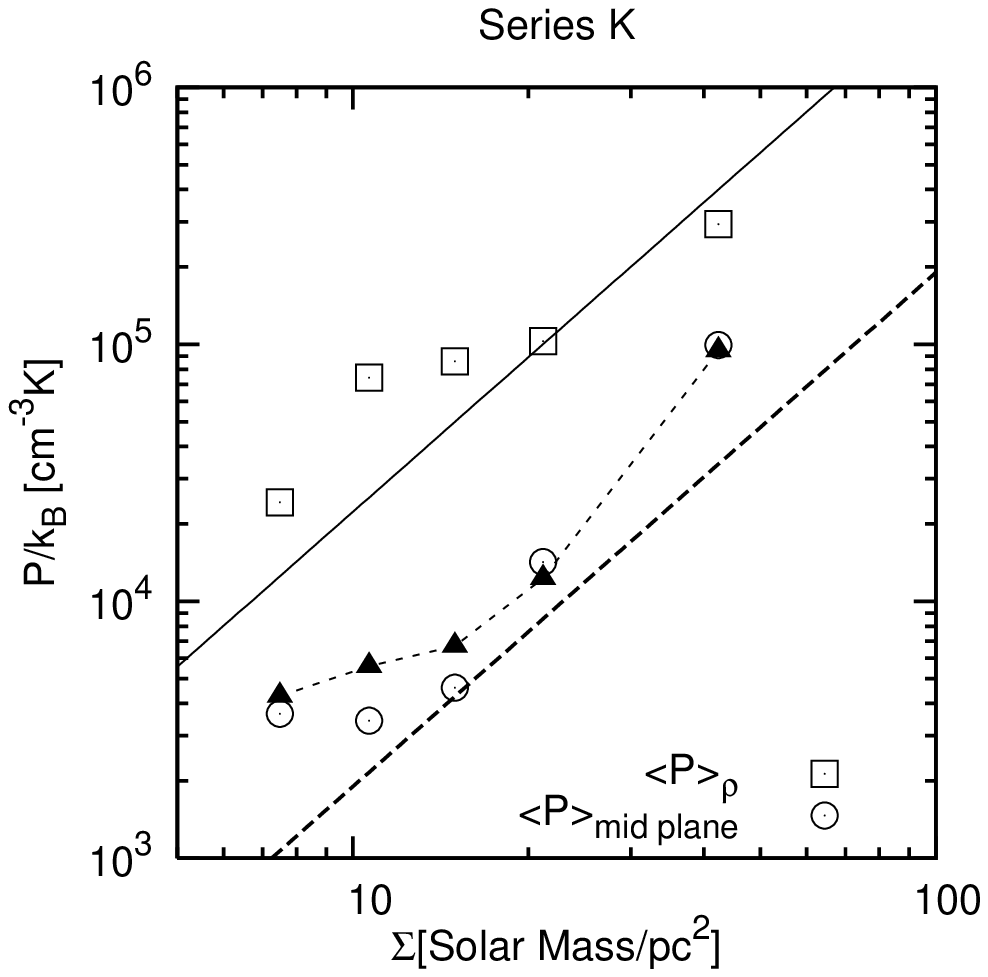}

\plottwo{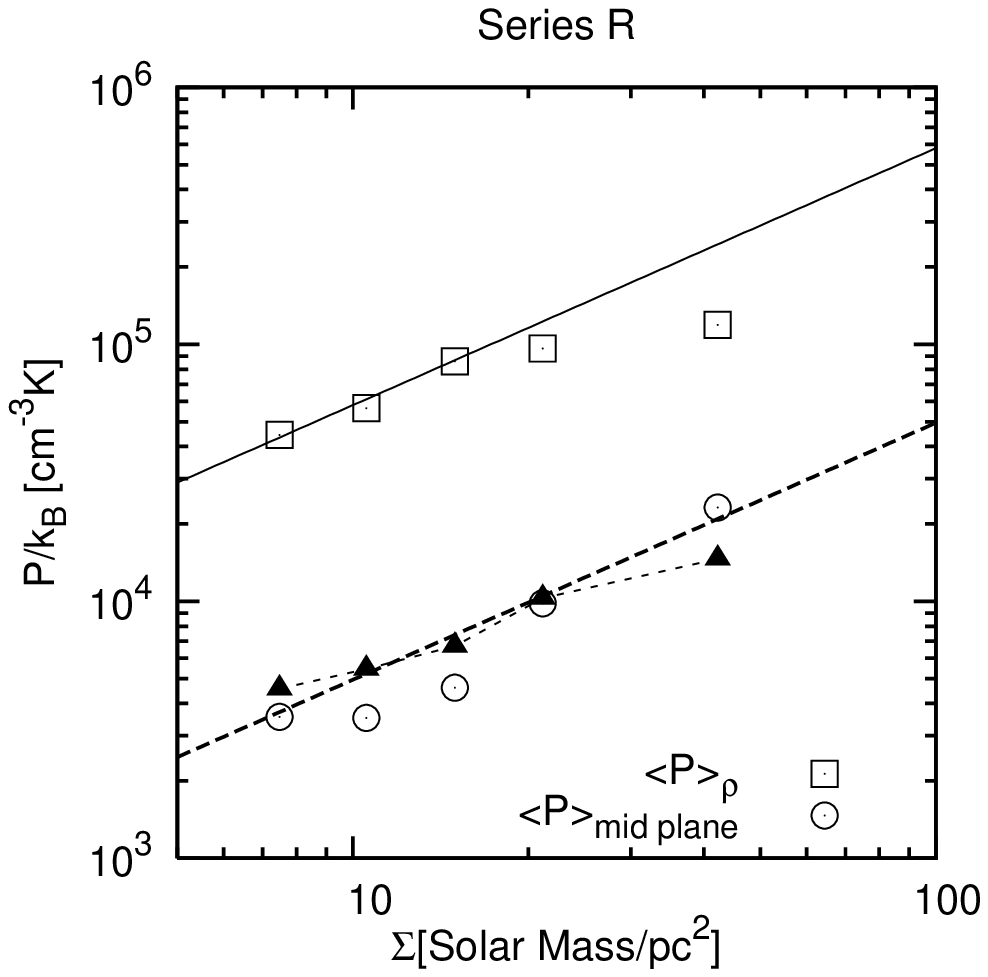}{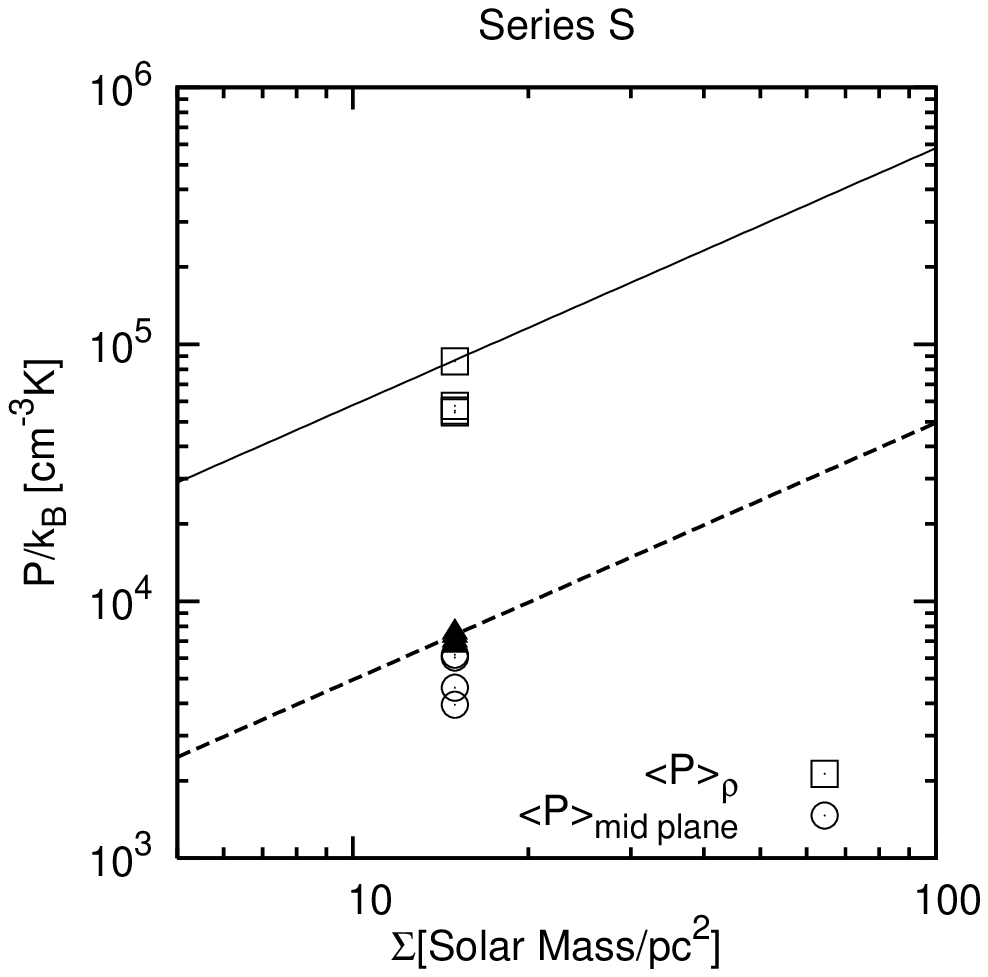}
 \caption{Measured, fitted, and estimated gas pressures. 
{\it Open boxes} and {\it circles} are the same as in Figure
 \ref{fig:Pressure}.
{\it Solid} and {\it dashed} lines are the corresponding fits to 
equation (\ref{Pfit_eq}). Our vertical-equilibrium midplane 
thermal pressure
estimate (eq. \ref{eq:Pnewth}) is plotted as {\it filled triangles}.}
\label{fig:Pressure2}
\end{figure}

In \S \ref{scale_height}, we defined an ``average'' vertical
equilibrium using the total surface density and the total vertical
velocity dispersion, and showed that this could yield an accurate
measurement of the disk thickness.  The same model (see Appendix) can
be used to estimate a midplane ``effective hydrostatic pressure,'' 
which we can compare to
measured values.  If $H$ is the scale height, then in equilibrium 
the mean midplane gas density is $\rho_0=\Sigma/(\sqrt{2\pi}H)$.
Using the total velocity dispersion, the predicted total 
gas pressure at the midplane is then given by 
$P_{0,{\rm tot}}=\sigma_z^2\rho_0$,
which using equation (\ref{Hest_eq}) gives 
\begin{eqnarray}
P_{0,{\rm tot}}&=&
\frac{\sigma_z^2\Sigma}{\sqrt{2\pi}H_{{\rm est},c+v_z}}
=\Sigma\left(G\Sigma+\left[(G\Sigma)^2+2G\rho_{\ast}\sigma_z^2\right]^{1/2}\right)\cr
&=&\Sigma \sigma_z \sqrt{2 G\rho_{\ast}}
(A+\sqrt{A^2+1}).
\label{eq:Pnew}
\end{eqnarray}
The expression (\ref{eq:Pnew}) corresponds to an extension of the
pressure estimate formula of BR06 using the inverse of the
$A$-dependent correction
factor that appears in the scale height estimate (\ref{Hest_eq}).
This correction factor is $>1$.

Equation (\ref{eq:Pnew}) gives an estimate of the total midplane
pressure, but the thermal pressure should represent only a fraction
$\langle c_s^2\rangle/\langle c_s^2 + v_z^2\rangle= 1- \langle
v_z^2\rangle/\sigma_z^2$ of $P_{0,tot}$, where $\langle v_z^2\rangle^{1/2}$
is the mass-weighted RMS turbulent velocity dispersion in the vertical
direction.  Thus, our estimate of the mean thermal pressure at the
midplane is
\begin{equation}
P_{0,{\rm th}}=\langle c_s^2\rangle \frac{\Sigma}{\sqrt{2\pi}H_{{\rm
    est},c+v_z}}
=\Sigma \frac{\langle c_s^2\rangle}{\sigma_z} \sqrt{2 G\rho_{\ast}}(A+\sqrt{A^2+1}).
\label{eq:Pnewth}
\end{equation}
In Figure \ref{fig:Pressure2}, we compare the pressure estimate from
equation (\ref{eq:Pnewth}) ({\it filled triangles}) with the
measurements of midplane pressure.  At large
$\Sigma$, the agreement is quite good, while at low $\Sigma$ the
estimated midplane pressures lie slightly above the measured values.
This behavior is similar to our results for estimated scale heights, 
which were in best agreement with the measured $H_{\rm ave}$ at large
$\Sigma$ (where the dense gas dominates the mass, and the velocity
dispersion is turbulence-dominated).

For all the Series in which $\Sigma$ is the independent variable,
we have fit the measured gas pressure to the formula:
\begin{eqnarray}
P/\kB
=D\sqrt{\frac{\rho_{\ast}}{\msun\pc^{-3}}}
\left(\frac{\Sigma}{\msun\pc^{-2}}\right).
\label{Pfit_eq}
\end{eqnarray} 
For $\langle P\rangle_{\rho}$ and $\langle P\rangle_{\rm midplane}$,
we find the respective coefficients are $D_{\rho}=1.3\times 10^4 \K
\cm^{-3}$ 
and
$D_{\rm midplane}=1.1\times 10^3 \K
\cm^{-3}$, respectively.  The largest and
smallest surface density models are excluded in the fits.  The results
of the fits are displayed as {\it solid} and {\it dashed} lines,
respectively, in Figure \ref{fig:Pressure2}.  To compare with the BR06
formula, we also fit $\langle P\rangle_{\rho}$ and $\langle
P\rangle_{\rm midplane}$ to $P=C \Sigma v \sqrt{2G\rho_{\ast}}$ with
$v=8$ km/s. We find $C_{\rho}=3.6$ and $C_{\rm midplane}=0.3$;
i.e. the BR06 formula for pressure yields values that are typically a
factor $\sim 3$ larger than our measured midplane pressures, and a
factor $\sim 4$ below the mass-weighted mean values of pressure. 
As noted above, the mass-weighted average pressures are
about ten times larger than the midplane pressures; this is evident in
the ratio of the fitting coefficients.

Finally, we note that for most models (except at low $\Sigma$), the measured
midplane pressure exceeds the maximum pressure of the warm neutral
medium, $P_{w, {\rm max}}/\kB= 5.5\times 10^3 \K \cm^{-3}$ 
for our adopted heating and cooling functions.
Dense clouds that are externally confined by the warm medium cannot
have pressure exceeding $P_{w, {\rm max}}$ unless they are internally
stratified (implying they are self-gravitating);  thus, 
$P_{w, {\rm max}}$ is the largest the midplane pressure could be in the absence of
self-gravity.  
Equation (\ref{eq:Pnewth}) can be solved for $\Sigma$ in terms of the
midplane value of $P_{0, {\rm  th}}$.  The maximum surface density for an
atomic-only 
disk without self-gravitating clouds is then obtained by setting 
$P_{0,{\rm  th}}\rightarrow P_{w, {\rm max}}$, with the result
$\Sigma \rightarrow (P_{w, {\rm max}}/G)^{1/2}\sigma_z/c_s$ times a
function of $A$ that varies between 0.3 and 0.6 for $A= 0.1 - 1$.
Assuming  $\sigma_z/c_s \sim \sqrt{2}$ and taking 
$P_{w, {\rm max}}/\kB=5.5 \times 10^3 \K \cm^{-3}$, the maximum
surface density for a pure-atomic disk is 
$\sim 10 \msun\pc^{-2}$; this is consistent with the saturation
levels for HI gas observed e.g. by \cite{2002ApJ...569..157W}.
Since the measured midplane pressure is a
volume-weighted sum of the pressures in different phases, a mean value
exceeding $P_{w, {\rm max}}$ implies that self-gravitating dense clouds occupy
a non-negligible fraction of the midplane volume,
$f_V=(M_{\rm dense}/M_{\rm diffuse})(\rho_{\rm diffuse}/\rho_{\rm dense})$, with
$\langle P\rangle_{\rm midplane}= (P_{\rm dense}- P_{\rm diffuse})f_V+
P_{\rm diffuse}$. 
In the next section, we turn to a discussion of the
relationship between the dense-to-diffuse mass ratio and global parameters.

\section{An Application: Molecular Mass/Pressure Relations}

In this section, we explore relationships between the dense gas
fraction and ``environmental'' conditions, including the gas pressure
and the gas surface density.  We are motivated by observations that
show high molecular fractions in environments -- including spiral arms
and galactic center regions -- where both the total gas surface
density and stellar density are high.  In particular, BR06 found for a
number of disk systems that the mean ratio of molecular-to-atomic mass
scales nearly linearly with the pressure estimate $P_{\rm BR}$ defined
in equation (\ref{PBR_eq}).  Although our turbulent, multiphase
simulations show that $P_{\rm BR}$ in fact overestimates the pressure
of the typical volume element and underestimates the pressure of the
typical mass element, $P_{\rm BR}$ nevertheless systematically
increases in a similar way to both $\langle P\rangle_{\rm midplane}$
and $\langle P \rangle_{\rho}$.  Thus, it is interesting to test how
the dense-to-diffuse gas mass ratio depends on the true values of
pressure.  In addition to empirical results suggesting a relation
between mass ratio and pressure, there are theoretical reasons that
the mass ratio should depend on the mean gaseous surface density.  For
example, if
atomic gas is converted to molecular clouds through gravitational
instabilities on a timescale $t_{\rm form}\sim
\sigma_{\rm HI}/(G\Sigma)$, and molecular clouds are destroyed by star
formation on a timescale $t_{\rm dest}$, equating cloud formation and
destruction rates implies $M_{\rm H_2}/M_{\rm HI}=t_{\rm dest}/t_{\rm form}$,
which is $\propto \Sigma$ if the HI velocity dispersion and cloud
destruction time are relatively constant.  Thus, it is interesting to
explore dependence of $M_{\rm H_2}/M_{\rm HI}$ on the surface density -- which
appears in both the effective hydrostatic pressure and the rate of
self-gravitating instabilities.

\subsection{Molecular Gas
\label{mol_fraction_disc}
}

\begin{figure}
\epsscale{0.6}
\plotone{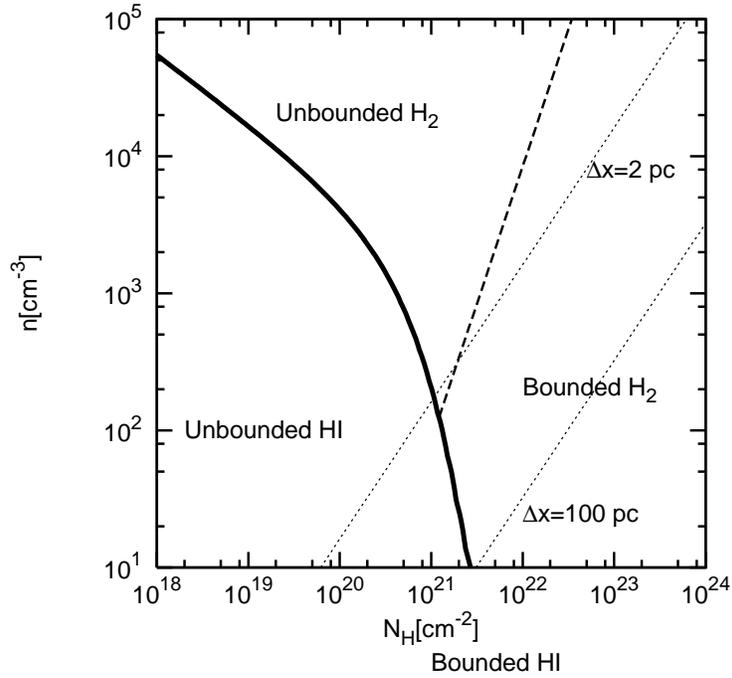}
 \caption{Phases in the density-column density plane. The 
{\it solid line} divides the area into predominantly-HI at low 
$n_{\rm H}$ and
$N_{\rm H}$, and  predominantly-H$_2$ at high $n_{\rm H}$ and $N_{\rm H}$, adopting
molecule formation and destruction processes as described in the text.  
Shown as a {\it dashed line} is $N=n L_{\rm J}$, where $L_{\rm J}$ is the
Jeans length at $n$ 
for $T=10$K gas.  The left
 and right hand sides of this line are gravitationally unbound and
 bound, respectively. Dotted lines show the typical resolution limit of the
 simulations $\Delta x=2$ pc, and a maximum cloud scale of $100$ pc.
}
\label{fig:Molecular}
\end{figure}

Although our numerical model does not directly include
formation/dissociation processes of H$_2$, we can nevertheless relate
our results to observed gas phases in an approximate way, using
density as a proxy.  Namely, we expect gravitationally
bound dense clouds at $n>100$ cm$^{-3}$ to consist primarily of H$_2$,
whereas diffuse gas at lower densities consists primarily of HI.
We argue for this approximate identification based on the
formation/dissociation equilibrium condition for 
$H_2$ molecules, which includes photodissociation and cosmic ray
dissociation, and formation on dust grains:
\begin{eqnarray}
(R_{\rm pump}&+&\zeta_{\rm CR}^{\rm H_2})nf_{\rm mol} 
= R_f n^2(1-f_{\rm mol})
\end{eqnarray}
\citep{1985ApJ...291..722T}.
Here, $f_{\rm mol}\equiv 2 n({\rm H}_2)/n$ is the molecular fraction, and
$1-f_{\rm mol}=n({\rm H I})/n$ is the atomic fraction.  
The FUV dissociation rate is limited by shielding, which depends on
the optical depth in H$_2$ lines and the extinction.  Formation on grains
depends on the sticking probability.  The details of the terms
involved are listed in Table \ref{tbl:param}.  We adopt 
FUV field strength $G_0=1.7$, 
gas and dust temperature $T=10$ K, and
cosmic-ray ionization rate of hydrogen atoms
$\zeta_{\rm CR}^{\rm H}=1.8\times 10^{-17}$s$^{-1}$.
For any total hydrogen column $N_{\rm H}$ and volume density $n$, we can
solve to obtain $f_{\rm mol}$, the molecular fraction.  
Figure \ref{fig:Molecular} shows, in the $n-N$ plane, the boundary
({\it solid line}) between the predominantly-atomic and
predominantly-molecular regimes, which we define by the locus of
points for which $f_{\rm mol}=0.5$.

\begin{table}
\begin{center}
\caption{Processes and parameters for H$_2$ formation/dissociation}
\label{tbl:param}
\begin{tabular}{lllll}
\hline\hline
Formation of H$_2$ on dust grains
&$R_f$&=& $6\times 10^{-17}(T/300)^{0.5}S(T)~{\rm cm^{3}s^{-1}}$ &
[1]\\
Sticking probability
&$S(T)$&=& $[1+0.04(T+T_d)^{0.5}+2\times 10^{-3}T +8\times
  10^{-6}T^2]^{-1}$&
[1]\\
Photo dissociation rate
&$R_{\rm pump}$ &=& $3.4\times 10^{-10}G_0\beta(\tau)\exp(-2.5A_v) ~ 
{\rm s^{-1}}$ & [1] \\
Self-shielding function
&$\beta(\tau)$ & & $\cdots$ & [1] \\ 
Optical depth
&$\tau$ &=& $1.2\times 10^{-14}fN_{\rm H}\delta v_d^{-1}$ & [1] \\
Cosmic-ray dissociation
&$\zeta_{\rm CR}^{\rm H_2}$&=& $2.29\zeta_{\rm CR}^{\rm H}$ & [2]\\
Turbulent line broadening
&$\delta v_d$ &=& 
$1~{\rm km/s}\left(\frac{N_{\rm H}/n}{1~\pc}\right)^{0.5}$ &
[3]
\\
Visual attenuation
&$A_v$ &$\equiv$& $N_{\rm H} /1.5\times 10^{21} \cm^{-2}$ &\\
\end{tabular}
\tablenotetext{}{[1] \cite{1985ApJ...291..722T}}
\tablenotetext{}{[2] \cite{1989ApJ...342..306H}}
\tablenotetext{}{[3] \cite{1987ApJ...319..730S}}

\end{center}
\end{table}

At any density, we can also define the Jeans length $L_{\rm J}=c_s
(\pi/G\rho)^{1/2}$, where $c_s^2=\kB T/\mu$ (we adopt $T=10$K).  
This defines a corresponding total column
of gas, $N_{\rm H}=n L_{\rm J}$, that could be expected to be gravitationally
bound.  The boundary between gravitationally unbound (low $n$ and $N$)
and bound (high $n$ and $N$) gas, based on this criterion, is shown in
Figure \ref{fig:Molecular} as a {\it dashed} line with $n\propto N^2$.
Note that if instead of $L_{\rm J}$ we had chosen as a length scale the
diameter $D$ of sphere containing mass equal to the Bonnor-Ebert
\citep{1956MNRAS.116..351B,1957ZA.....42..263E} critical mass,
$M_{\rm BE}=1.182 c_s^3 /(G^3\rho)^{1/2}$, then $D=0.74 L_{\rm J}$. This would
shift the unbound/bound line in the $\log(N)-\log(n)$ plane to the
left by $\log(0.74)=-0.13$.  

We note that the gravitational binding criterion discussed above considers only
support by thermal pressure.  Turbulence can lend further support
against gravity, and this is particularly important for molecular gas,
which is quite cold.  For example, if we considered
turbulence-supported clouds with velocity dispersion following the
observed linewidth-size relation of Galactic GMCs
\citep{1987ApJ...319..730S}, then the column density separating
gravitationally bound from unbound regions would have a constant value
equal to half of the mean observed GMC column, amounting to $N_{\rm H}=7.5
\times 10^{21} \cm^{-2}$.  A higher normalization for the
linewidth-size relation (as occurs in galactic center regions; see
\citealt{2001ApJ...562..348O}) would further shift the unbound/bound
limit to larger $N$.  Thus, moderate-density molecular gas can in
principle be gravitationally unbound under conditions of sufficiently high
turbulence \citep{1993ApJ...411..170E}.  For our current simulations,
however, turbulence levels are not this high (see discussion below).

We have denoted the three different regions in the $\log(N)-\log(n)$
plane according to their expected chemical and gravitational
properties.  The crossing point of the two separation loci is at
$n\approx 125$ cm$^{-3}$ and 
$N_{\rm H}\approx 1.2\times 10^{21} \cm^{-2}$
(Av $\sim$ 0.75 mag), with corresponding local Jeans length of 
$L_{\rm J}=3.1$ pc.  
This size is in fact slightly larger than
the typical resolution limit of our simulations, $\Delta x =2$pc; we show this
limit in Figure \ref{fig:Molecular} as a {\it dotted} line, with
regions to the right resolved and those to the left below the
resolution limit.  The resolution limit crosses the HI/H$_2$ separation
curve at 
$n=180~{\rm cm}^{-3}$.
Because the resolution limit falls at larger $N$ than the bound/unbound
separation nearly everywhere in the molecular domain, all zones at a given
density that are resolved and molecular would also be gravitationally
bound.  In practice, clouds do not exceed $\sim 100$ pc in crossection; we have
marked this limit in the Figure as a {\it dotted} line.

According to the limits shown in Figure \ref{fig:Molecular}, any resolved
regions in our simulations at $n>100\cm^{-3}$ would be molecular.
This is a conservative definition, since it omits some gas between
$n\sim 10-100$ with  $N_{\rm H} > 10^{21} \cm^{-2}$
that could be molecular.  However, gas at these
densities could also be in the cold atomic phase (which extends down
to $n_{\rm cold, min}= 8.6 ~\cm^{-3}$ for the cooling curve we
adopt); we choose
the stricter definition.  We note that when the virial ratio
($\sim$kinetic/gravitational energy; see \citeauthor{Paper1}) is
measured for gas in the range $n=1-100\cm^{-3}$ (most of which is at
$10\cm^{-3}<n$), the values are well above unity -- implying that gas
parcels in this density range are mostly found in non-self-gravitating
regions with low surrounding
column densities, to the left of the unbound/bound curve.  
This suggests that in practice very little
high-column density gas that would be molecular is missed when we set
the minimum threshold at $100\cm^{-3}$.  From the point of view of
dynamics, this is because the density rises whenever any
region becomes gravitationally bound, so low-density  regions at
high column are rapidly depopulated.  
We also note that the H$_2$
formation time in dense, cold regions is expected to be short ($\sim
10^6$ years from \citealt{2007ApJ...659.1317G}), because supersonic
shocks increase the density above ambient values 
and accelerate the molecule formation process, which occurs at a rate
$n R_f$.

\subsection{Molecular Mass-Pressure Relation}

\begin{figure}
\epsscale{1.0}
\plottwo{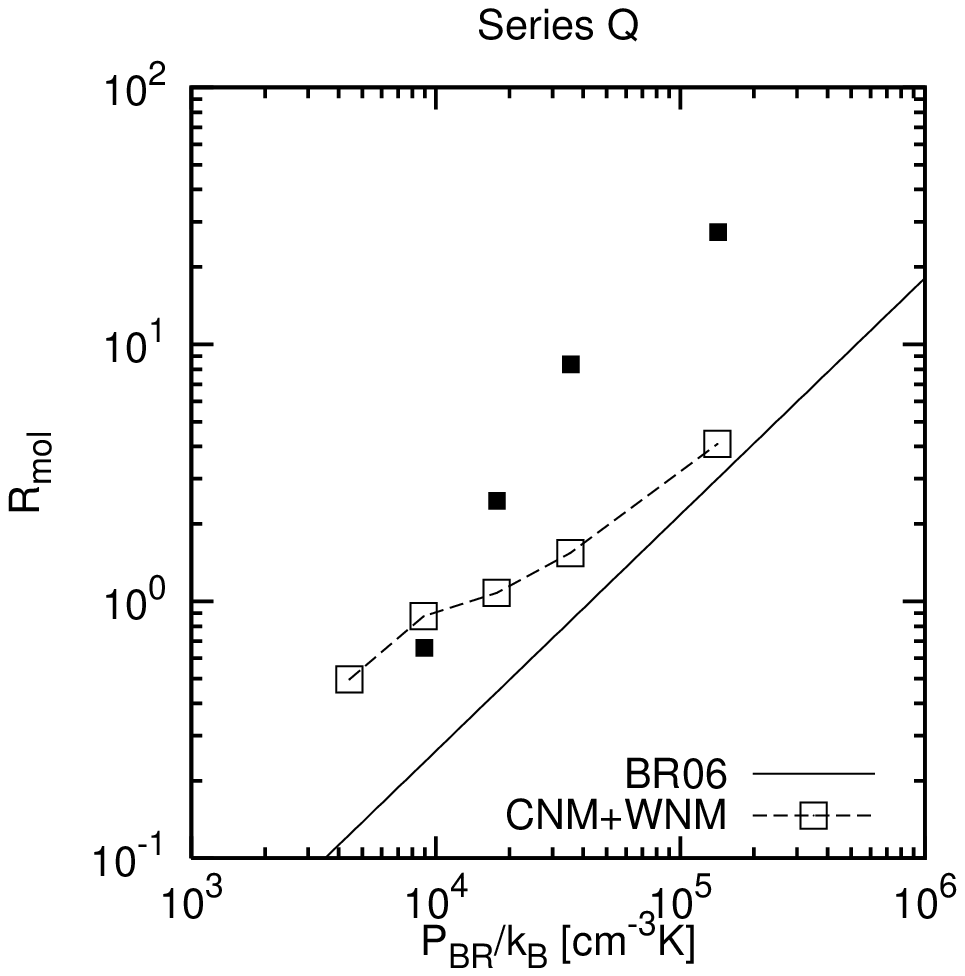}{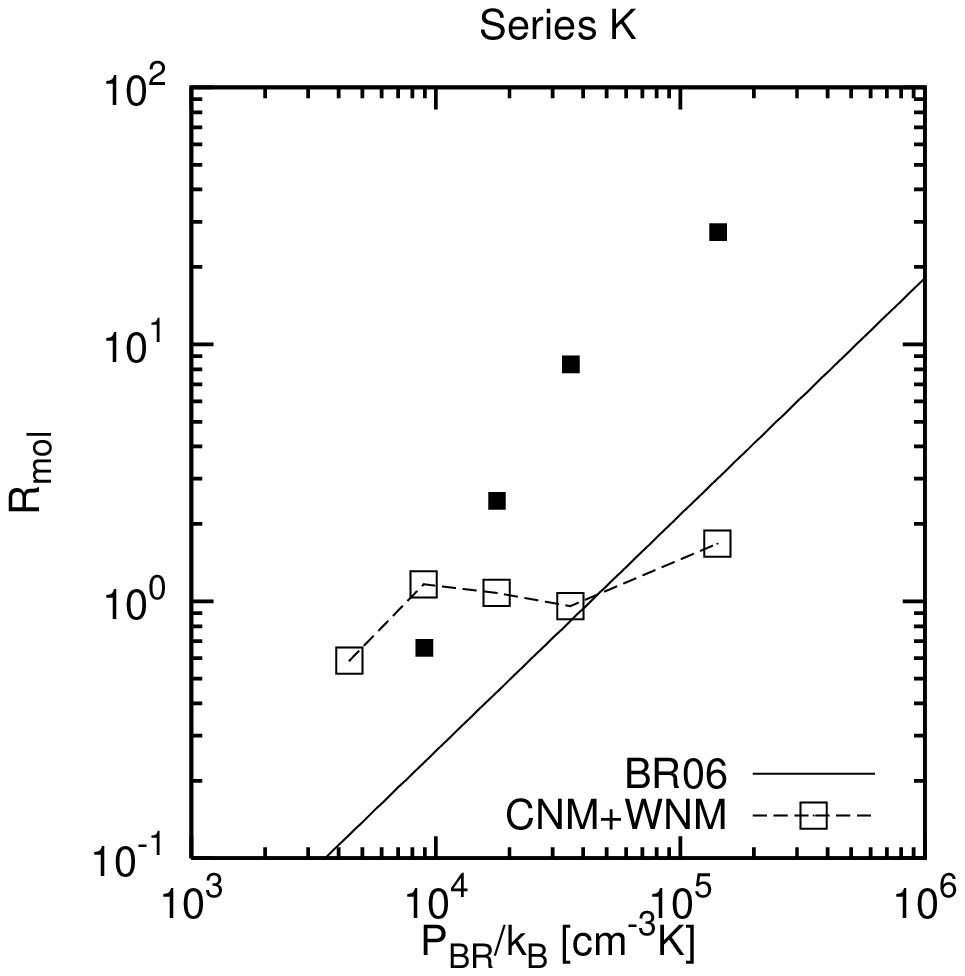}

\plottwo{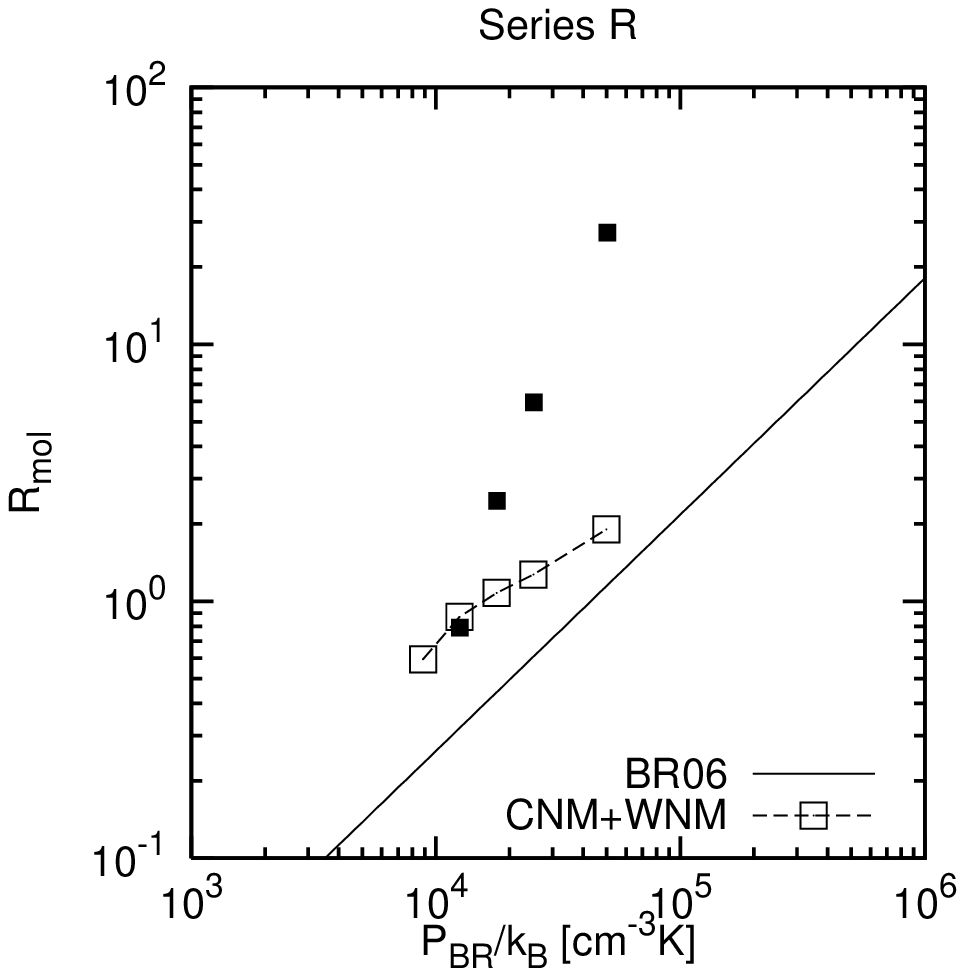}{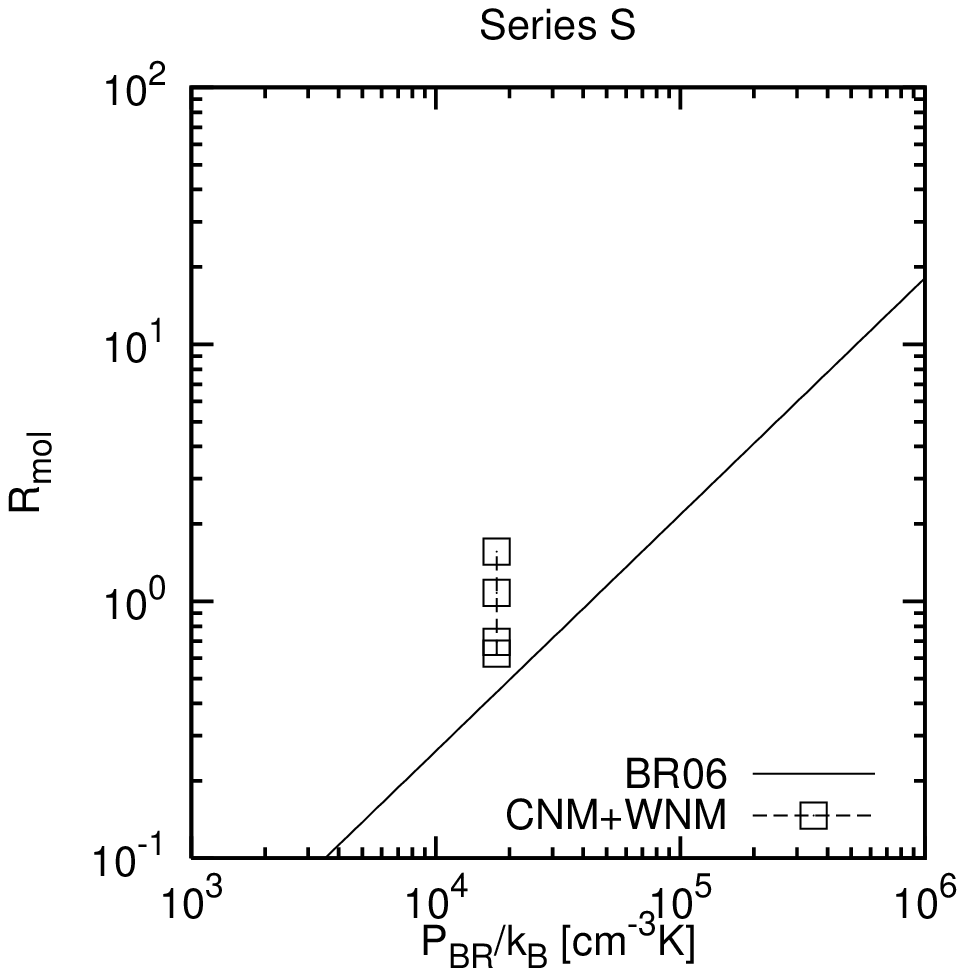}
 \caption{Mean molecular-to-atomic mass ratio $R_{\rm mol}$ as a
   function of $P_{\rm BR}$ (see eq. \ref{PBR_eq}), shown as {\it open
   boxes} for all hydrodynamic models.  {\it Filled boxes} show the
   results from hydrostatic models.  The {\it solid line} shows the
   empirical fit from BR06 (eq. \ref{eq:Rmol}).}
\label{fig:H2P}
\end{figure}

Following the discussion in the previous section, we adopt a working
definition of molecular gas as that at $n_{\rm H}>100 \cm^{-3}$.  Atomic gas
therefore consists of the lower-density complement, including both
what would be observable as warm and as cold HI in 21 cm emission.
The mass ratio of molecular to atomic hydrogen is
 then defined as
\begin{eqnarray}
R_{\rm mol}\equiv 
\frac{M(n>100\, {\rm cm}^{-3})}{M(n<100\, {\rm cm}^{-3})},
\end{eqnarray}
where we apply space- and time-averages before taking the ratio.

Figure \ref{fig:H2P} shows $R_{\rm mol}$ as a function of 
$P_{\rm BR}=
\Sigma v \sqrt{2G\rho_{\ast}}$ for all
hydrodynamic and hydrostatic Series (we use $v=8\kms$ as in BR06).  
We also show as a {\it solid
  line} the empirical fitting formula from the observational study of
BR06 (see their eq.13):
\begin{equation}
R_{\rm mol}=
\left[\frac{P_{\rm BR}/\kB}{4.3\times 10^4 \K \cm^{-3}} \right]^{0.92}. 
\label{eq:Rmol}
\end{equation}
Interestingly, we find that our results for $R_{\rm mol}$ follow the
empirical result for some but not all series.  In particular, the
models in Series Q and R -- which have values of $\Omega$ that scale
with $\Sigma$ in such a way as to keep the gaseous Toomre parameter
constant -- are close to the BR06 fit.  The models in Series K, which
have constant $\kappa$ and therefore high (or low) values of
$\kappa/\Sigma$ where $P_{\rm BR}$ is low (or high, respectively), do
not follow the empirical result of BR06, but instead show a ratio
$R_{\rm mol}$ that is near unity independent of $P_{\rm BR}$.  This
has two interesting implications. First, our models with $\Omega
\propto \Sigma$ have similar behavior to real galaxies, indicating
that real systems evolve (by converting their gas to stars) in such a
way as to have Toomre parameter fall within a limited range of values.
Second, because the K Series departs from the BR06 result, our models
suggest that the molecular fraction does not have a one-to-one
relationship to the effective pressure parameter $P_{\rm BR}$.  
Comparing series Q and K which have the same $\Sigma$ and
$\rho_{\ast}$, $R_{\rm mol}$ increases with increasing $\kappa$.  For
example, Figure 8 shows that $R_{\rm mol}$ increases by factor 2.4
when $\Omega$ (and $\kappa$) increases by factor $2\sqrt{2}$, for the
highest-$\Sigma$ ($\Sigma= 42 \msun \pc^{-2}$) model. For the 
$\Sigma= 21 \msun \pc^{-2}$ model, the $R_{\rm mol}$ increase is 60\% 
for an $\Omega$ increase by a factor  $\sqrt{2}$, comparing the Q and
K series. 
Series S, which varies $\kappa$ at a given value of
$\Sigma$ and $\rho_{\ast}$, also shows departures from the empirical 
$R_{\rm mol}$ vs. $P_{\rm BR}$ relation. 

Given
that molecular gas in our models is primarily found in
gravitationally-bound systems, it in fact makes sense that the
molecular fraction should not have a one-to-one relationship to the
parameter $P_{\rm BR}$, since $P_{\rm BR}$ does not include any
effects of galactic rotation.  Galactic rotation and shear are crucial
for regulating the large-scale gravitational instabilities that create
giant molecular clouds in real galaxies as well as in our models, so
we believe that the molecular-to-atomic ratio must intrinsically be
sensitive to environmental factors that are not captured in $P_{\rm
BR}$.  Thus, if observed galaxies {\it do} show a one-to-one relation between 
$R_{\rm mol}$ and $P_{\rm BR}$, it implies that the environmental
parameters $\kappa$, $\Sigma$, and $\rho_{\ast}$ are not all independent in
real systems. 

\begin{figure}
\epsscale{1.0}
\plottwo{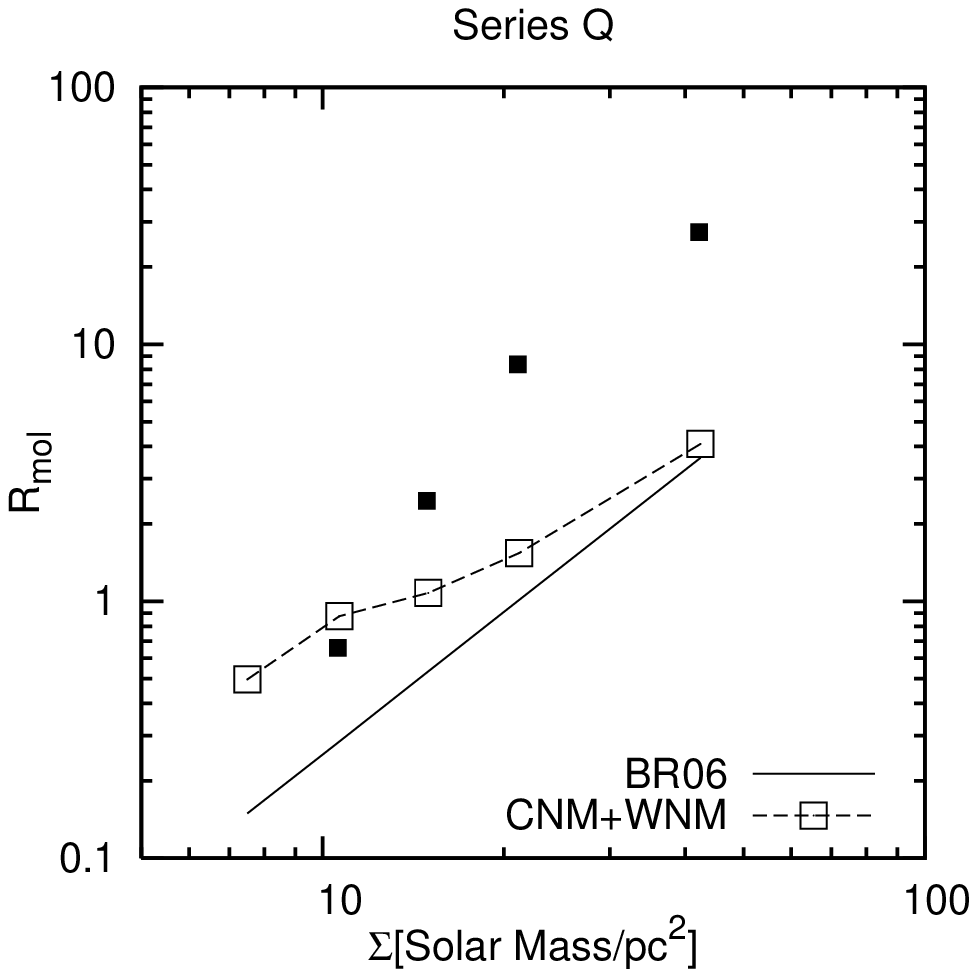}{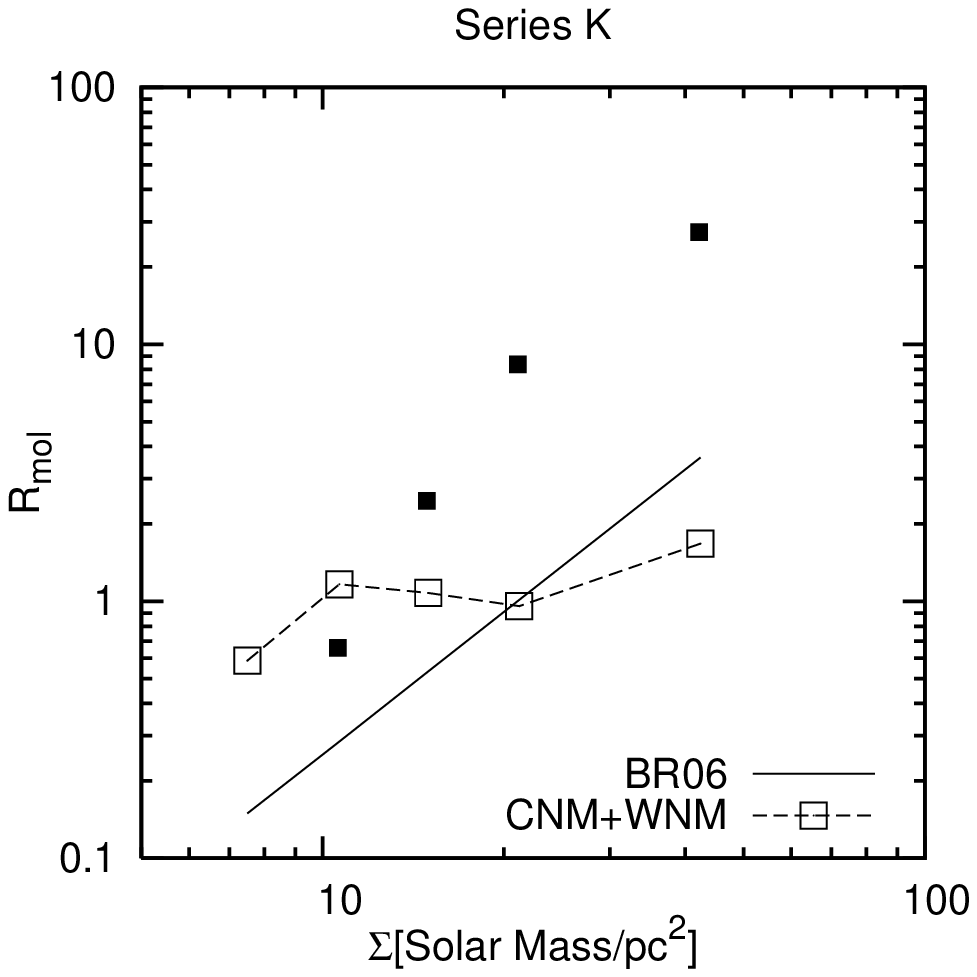}

\plottwo{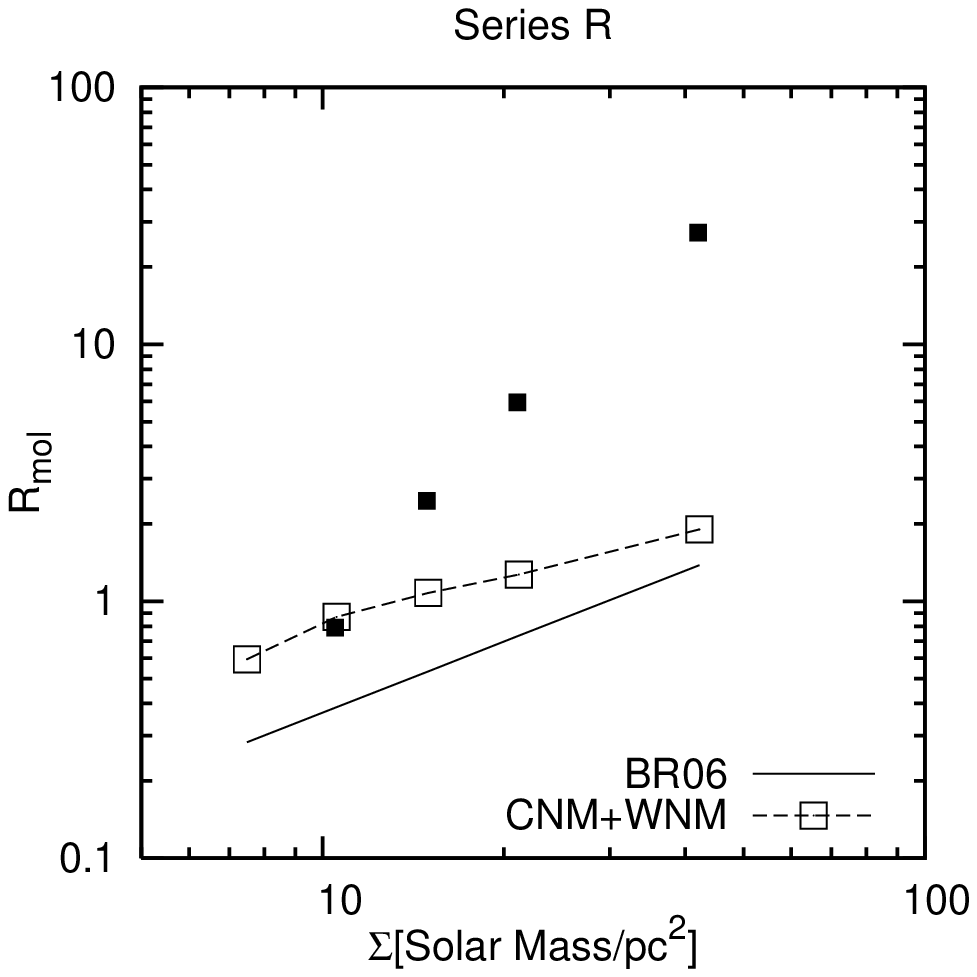}{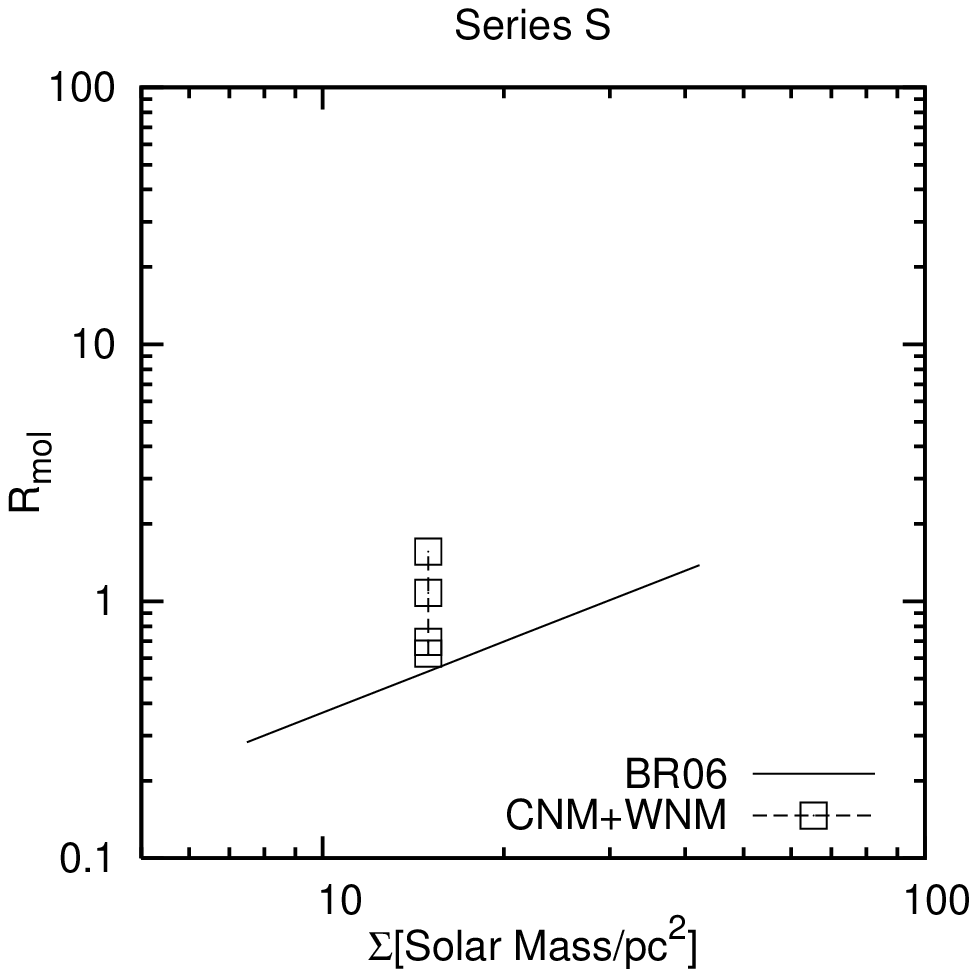}
 \caption{
Molecular-to-atomic mass ratio $R_{\rm mol}$ vs.
total gaseous surface density $\Sigma$. {\it Open boxes} show the
results from hydrodynamic models, and {\it filled boxes} show the 
hydrostatic model results (HSP for Series K and Q,
HSC for Series R). The {\it solid line} indicates the empirical result
from BR (eq. \ref{eq:Rmol} using eq. \ref{PBR_eq}).}
\label{fig:H2S}
\end{figure}

In Figure \ref{fig:H2S}, we show $R_{\rm mol}$ as a function of the
surface density for all of our model Series.  The behavior is similar
to that shown in Figure \ref{fig:H2P} because $P_{\rm BR}$ depends
monotonically on $\Sigma$ for all our Series: $P_{\rm BR}\propto
\Sigma^2$ for Series Q and K (which have $\rho_{\ast}\propto \Sigma^2$),
while $P_{\rm BR}\propto \Sigma$ for Series R (which has
$\rho_{\ast}=const.$).  From both Figures Figure \ref{fig:H2P} and
\ref{fig:H2S}, it is evident that the hydrostatic models ({\it filled
boxes}) generally have a much larger molecular component than both the
hydrodynamic models and the empirical results, except at low gaseous
surface density.  This indicates that turbulence is essential for
determining the phase balance between diffuse and dense gas in the ISM
as a whole.  If the ISM were a static system, it would be
overwhelmingly molecular even at fairly moderate values of $\Sigma$
and $\rho_{\ast}$.  In real galaxies, turbulence limits gaseous settling
into the midplane and the extreme self-compression that would
otherwise ensue.

\begin{figure}
\epsscale{1.0}
\plottwo{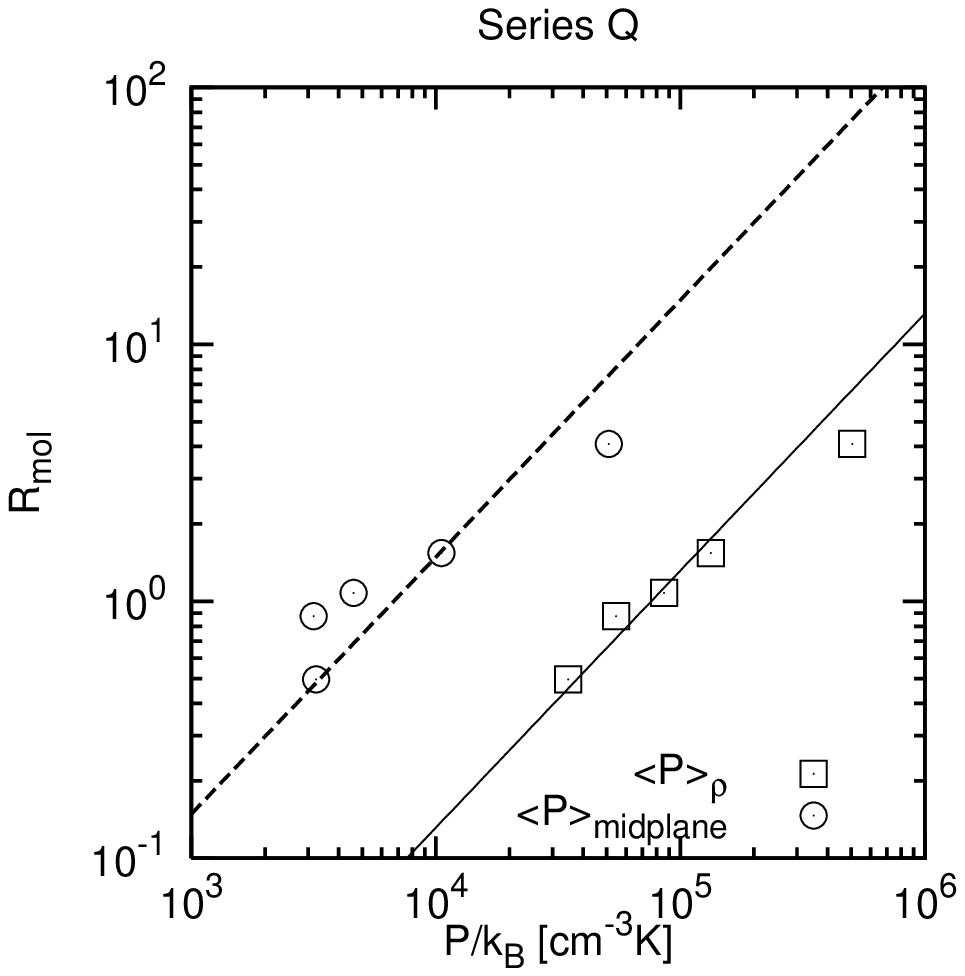}{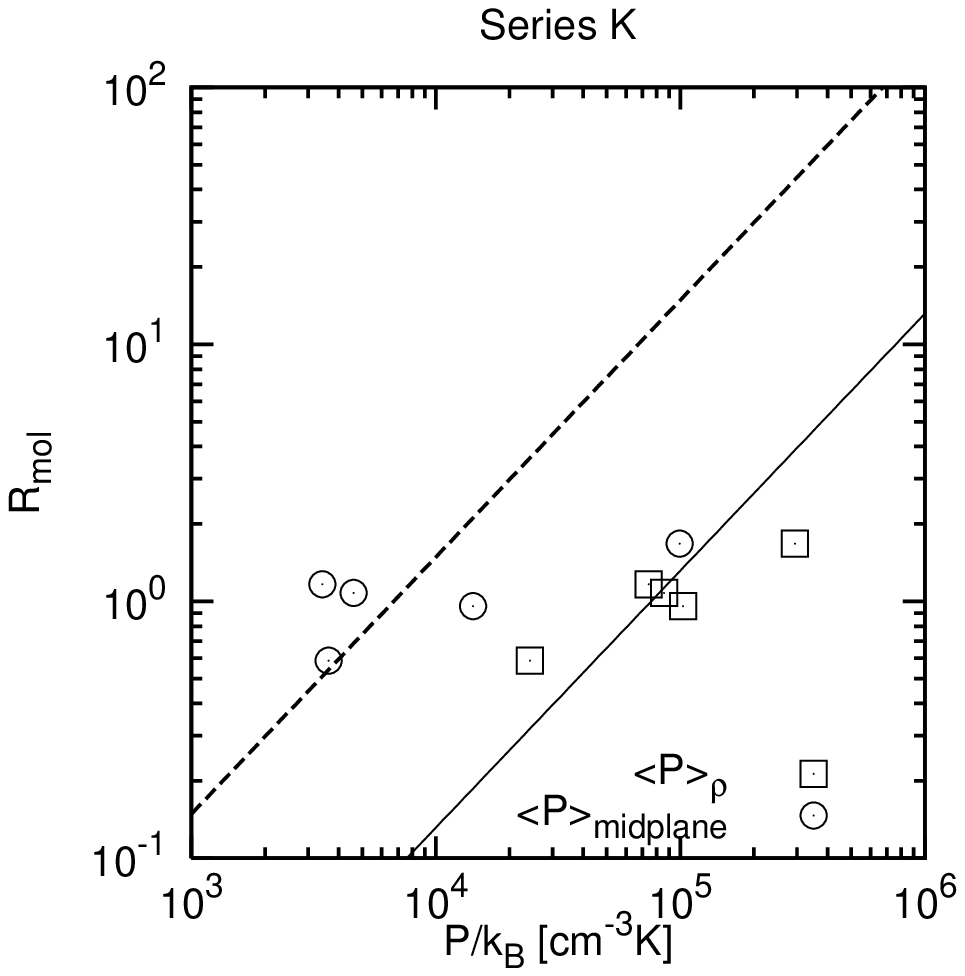}

\plottwo{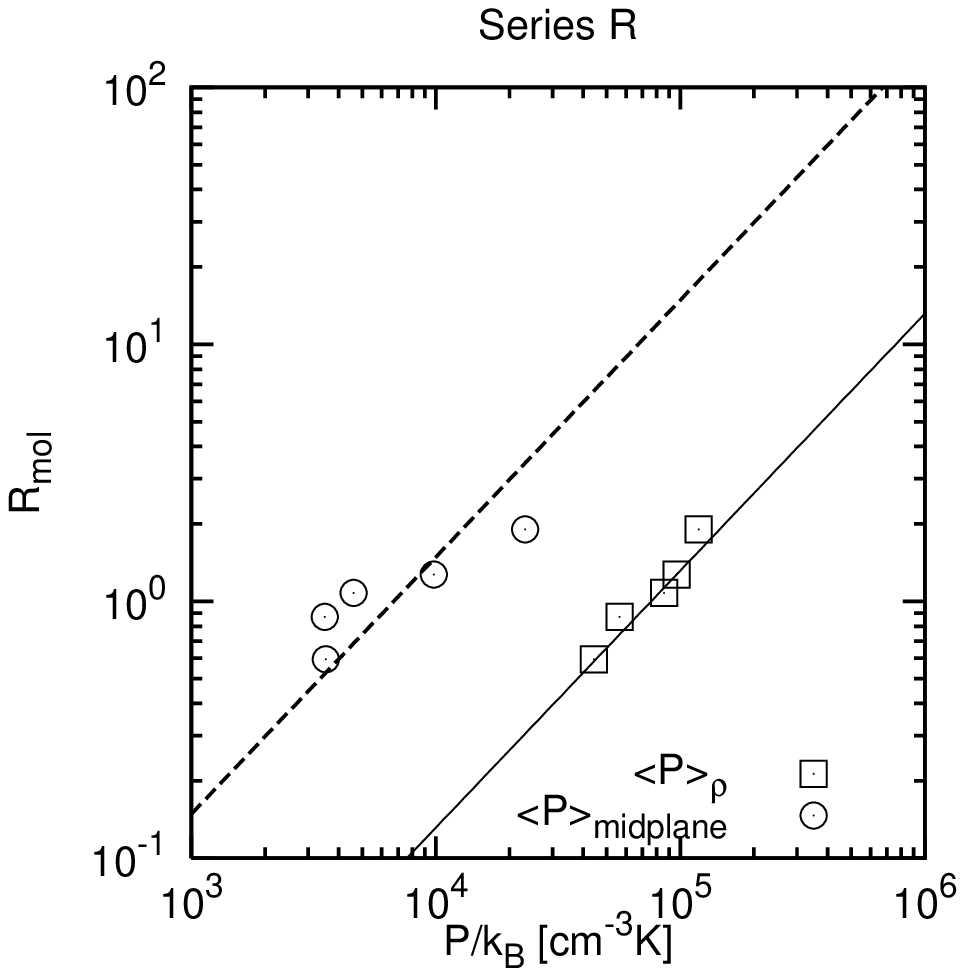}{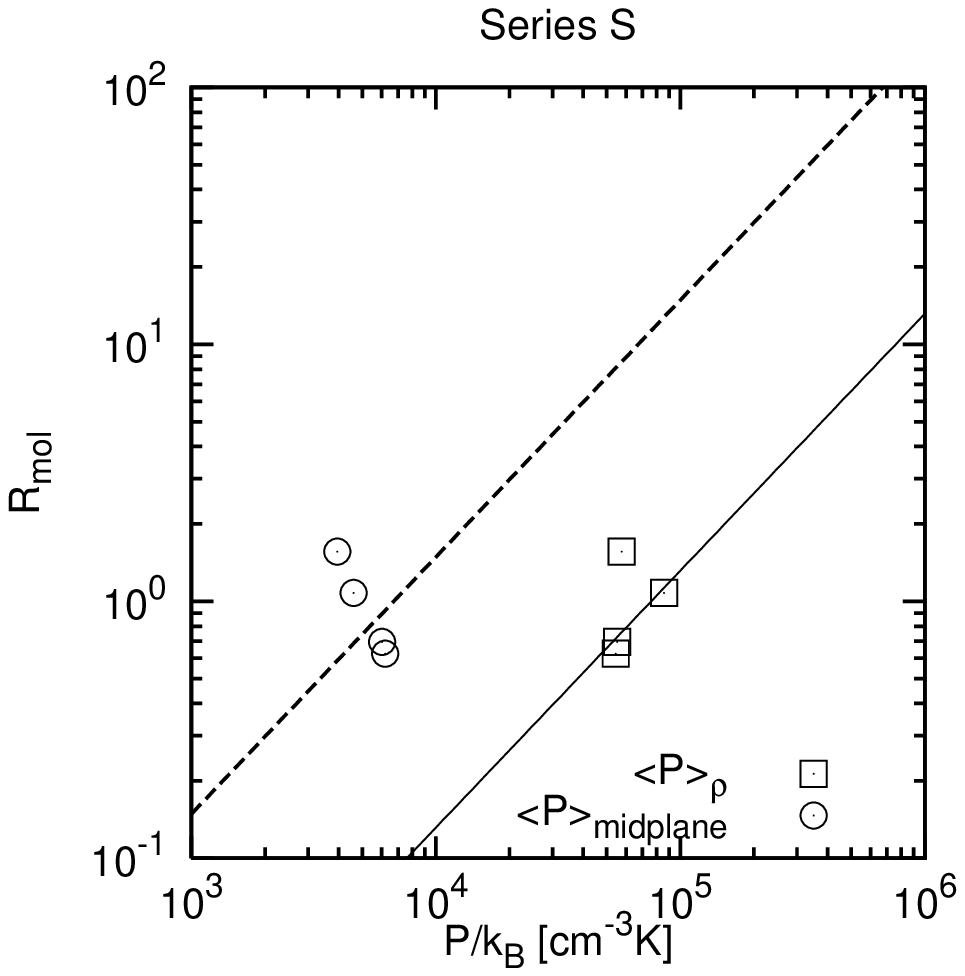}
 \caption{Molecular-to-atomic mass ratio $R_{\rm mol}$ vs. measured
   mass-weighted mean pressure $\langle P\rangle_\rho$ 
   and midplane pressure $\langle P\rangle_{\rm midplane}$ for all  
 hydrodynamic models.  Lines show linear fits (see text).}
\label{fig:H2Pressure}
\end{figure}

While we have argued that the molecular content of a galaxy cannot (in
general) be predicted solely from $\Sigma$ and $\rho_{\ast}$ because
self-gravitating horizontal contraction is also responsive to the
local rotation and shear rates, it still is plausible that the
molecular fraction should reflect the {\it true} mean pressure in the
ISM.  If molecular gas is collected in self-gravitating clouds, then
since their internal pressure is higher than ambient levels, an
increase in the molecular fraction should go hand-in-hand with a
higher mass-weighted mean pressure $\langle P\rangle_\rho$.  At the
same time, ambient midplane pressures $\langle P\rangle_{\rm
midplane}$ increase when the total gas surface density increases, and
(provided that $\kappa$ is low enough)
larger $\Sigma$ also renders the disk susceptible to gravitational
instabilities 
that would form dense, bound clouds and increase the
molecular fraction.  

We explore these ideas by plotting in Figure
\ref{fig:H2Pressure} the molecular-to-atomic ratio against our two
measures of mean gas pressure, $\langle P\rangle_\rho$ and $\langle
P\rangle_{\rm midplane}$.  We also fit the combined results for all
Series to single linear relations. These fits, overplotted in Figure
\ref{fig:H2Pressure} as {\it solid} and {\it dotted} lines, are
$R_{\rm mol}=\langle P\rangle_\rho/[7.6 \times 10^4 \cm^{-3} \K \kB]$ 
and $R_{\rm mol}=\langle P\rangle_{\rm midplane}/[6.7 \times 10^3
\cm^{-3}\K \kB]$.  
For Series Q and R, the fits using $\langle P\rangle_\rho$ are quite
good, and the fits using $\langle P\rangle_{\rm midplane}$ are also
fairly good (Series Q and R also show better agreement with empirical
results than the other Series).  For Series K, the fit using $\langle
P\rangle_\rho$ is reasonably close to the models results, but the fit
based on $\langle P\rangle_{\rm midplane}$ fails in a similar manner
to that shown in Figure (\ref{fig:H2P}) and discussed above.  The
basic reason for this is that the midplane pressure, either directly
measured or estimated using equation (\ref{eq:Pnewth}), increases with
increasing $\Sigma$. However, the molecular-to-atomic ratio for Series
K does not strongly and secularly increase with $\Sigma$ due to the
differences in rotational effects in this constant-$\kappa$ Series
compared to the other Series.  At high values of $\Sigma$ in Series K,
the disk is extremely gravitationally unstable overall, and as a
consequence is more active in producing feedback than other models at
the same $\Sigma$.  As a consequence, a smaller fraction of the gas
mass ends up being in the dense phase than in Series Q and R.
Overall, we conclude that $R_{\rm mol}$ is indeed well correlated with
the mass-weighted mean pressure, $\langle P\rangle_\rho$, as (almost
definitionally) is expected.  The measured mean midplane pressure, which is 
more closely related to simple vertical-equilibrium
pressure estimates, is less well correlated with $R_{\rm
mol}$ when environmental parameters $\kappa$, $\rho_{\ast}$, and $\Sigma$
are all independent.

\section{Summary and Discussion}

We have used numerical simulations of turbulent, multiphase,
self-gravitating gas orbiting in the disks of model galaxies to study
the relationships among pressure, the vertical distribution of gas,
and the relative proportions of dense and diffuse gas.  In particular,
we compare the results on vertical stratification obtained from
space-time averages of fully-dynamic -- and often turbulence-dominated
-- systems with simple estimates based on single-component effective 
hydrostatic
equilibria.  We also investigate how vertical-equilibrium estimates
for the pressure compare with measured mean values of the pressure in
our models.  Empirical studies by BR06 have identified a linear
relation between the molecular-to-atomic mass ratio $R_{\rm mol}$ and
a midplane ISM pressure estimate, $P_{\rm BR} \propto \Sigma \sqrt{
\rho_{\ast}}$.  We study the origin and implications of this relation by
testing the correlations among $R_{\rm mol}$, $P_{\rm BR}$, and the
directly-measured midplane and mean pressures in our models.

Our chief conclusions, and their implications, are as follows:

1. The average disk scale height is well represented by estimates that
assume hydrostatic equilibrium and an effective total pressure based
on the total (thermal + turbulent) vertical velocity dispersion (see
Fig. \ref{fig:Have} and eq. \ref{Hest_eq}).  Thus, provided that gas
surface densities, vertical velocity dispersions, and stellar density
can be measured, an accurate estimate for the disk thickness can be
obtained.

Hydrostatic equilibrium with an effective turbulent pressure is
commonly assumed in both Galactic and extragalactic observational
studies (e.g.
\citealt{1991ApJ...382..182L,1994ApJ...433..687M,1995ApJ...448..138M,
1997A&A...326..554C,2000MNRAS.311..361O,
2002A&A...394...89N,2004ApJ...608..189D,
2004ApJ...612L..29B,2006ApJ...650..933B,
2008AstL...34..152K}), but to our knowledge the relations that are
generally adopted have not previously been tested with direct
numerical simulations.  Our hydrodynamic studies demonstrate that for
determining the scale height $H$, the effective hydrostatic
equilibrium assumption is indeed sufficient, even when turbulent
support far exceeds thermal support (and provided that magnetic
effects are sub-dominant; see below).  Thus, measured disk thicknesses
in edge-on disk galaxies could in principle be used to determine the
unobservable vertical velocity dispersion, and measured line-of-sight
velocity dispersions in face-on galaxies could be used to determine
the unobservable disk thickness.

The basic reason the hydrostatic formula can be
used to obtain an accurate measure of $H$ is that what is
really being equated is the total vertical momentum flux 
$\rho (\kB T/\mu + v_z^2)$ averaged over the midplane, and the total
vertical weight of the ISM, $\int dz\, \rho\, g_z \sim \rho 4 \pi G (\rho
+ \rho_{\ast}) H^2$, averaged over the horizontal direction.
Provided that the time-averaged value of the momentum per unit volume
in the midplane does not change, momentum conservation including
gravitational source terms demands that the difference between 
vertical momentum flux and vertical weight must be zero, 
independent of details of the dynamics.  The formula
$H^2\approx \sigma_z^2/[4\pi G (\rho_{\ast} + \Sigma/H\sqrt{2\pi})]$ is
therefore fundamentally an expression of momentum conservation.

2. Mass-weighted mean pressures $\langle P \rangle_{\rho}$ in our
hydrodynamic models significantly differ from the mean midplane
pressure $\langle P\rangle_{\rm midplane}$, while these quantities
are quite similar to each other in our static comparison models.
Typically, the hydrodynamic models yield values of $\langle P
\rangle_{\rho}$ an order of magnitude larger than $\langle
P\rangle_{\rm midplane}$. The difference can be attributed to
self-gravitating condensation, which makes concentrated clouds 
with high internal pressure
rather
than a horizontally-uniform gas distribution
with more moderate pressure.

Simple estimates of the pressure based on vertical hydrostatic
equilibrium fall between mass-weighted and midplane values, with the
formula used by BR06 (see our eq. \ref{PBR_eq}) comparable to the
geometric mean $P_{\rm BR} \sim \sqrt{\langle P \rangle_{\rho} \langle
P\rangle_{\rm midplane}}$. A single-component estimate for the
midplane thermal pressure that accounts for self-gravity and the mean
thermal and turbulent velocity dispersions (see eq. \ref{eq:Pnewth})
follows the measured midplane pressure fairly closely, especially at
high $\Sigma$.  Thus, if turbulent and thermal vertical velocity
dispersions can be measured directly (for face-on galaxies), a good
estimate of the midplane total or thermal pressure can be computed via
equation (\ref{eq:Pnew}) or (\ref{eq:Pnewth}).  For an edge-on system
in which the scale height is measured, the midplane total or thermal
pressure can be estimated as $\Sigma/(H \sqrt{2\pi})\times \langle
\sigma_z^2\rangle \ {\rm or}\ \langle c_s^2\rangle$.  Midplane
pressure estimates based on large-scale observables that assume
hydrostatic equilibrium can be quite accurate, but this depends on an
accurate measure of the vertical velocity dispersion or vertical
thickness.  Even if the velocity dispersion is not known, the {\it
relative} midplane pressures of different regions within a galaxy (or
from one galaxy to another) can be obtained using the hydrostatic
formulae, provided the variation in the (unknown) velocity dispersion
within the observational sample is small compared to the variation in
the stellar volume and gaseous surface densities.  Midplane pressure
estimates made in this way should not, however, be treated as a proxy
for the pressure in the typical mass element, $\langle P
\rangle_{\rho}$, which can be much larger than the pressure in the
typical volume element.

3. Based on calculations of molecular abundance as a function of
hydrogen volume density $n$ and column density $N$ combined with the
resolution and measured virial ratios in our simulations, we adopt a
working definition of gas at $n\ge 100 \cm^{-3}$ as ``molecular'' and
$n<100 \cm^{-3}$ as ``atomic''.  We then investigate the ratio $R_{\rm
  mol}=M_{\rm H_2}/M_{\rm HI}$ for all our models.  We find that 
Series Q and R, which have rotation rate $\Omega \propto
\Sigma$, show correlations between $R_{\rm mol}$ and $P_{\rm BR}$
(or $R_{\rm mol}$ and $\Sigma$) that are similar to the empirical 
result reported
by BR06, $R_{\rm mol}\propto P_{\rm BR} \propto
\Sigma\sqrt{\rho_{\ast}}$. On the other hand, Series K and S, in which $\Sigma$ and
$\Omega$ do not vary together, depart from the empirical relation
$R_{\rm mol} \propto P_{\rm BR}$.

We conclude that (i) the molecule fraction inherently depends on the
rotational state of a galactic disk, not just on the local values of
the stellar volume and gaseous surface densities $\rho_{\ast}$ and
$\Sigma$; and (ii) the empirical relation $R_{\rm mol} \propto P_{\rm
BR}$ identified by BR06 implies that the third ``environmental
parameter,'' the epicyclic frequency $\kappa = \sqrt{2} \Omega$
(assuming a flat rotation curve), is not independent of $\rho_{\ast}$ and
$\Sigma$ in real galaxies.  This dependence can be accomplished by
evolution: for example, disk galaxies may convert gas into stars until
they reach a state in which the Toomre parameter $\propto
\kappa/\Sigma$ approaches a critical value.

4. We have tested the correlation between $R_{\rm mol}$ and the
measured pressures in our models, $\langle P \rangle_{\rm midplane}$
and $\langle P \rangle_\rho$, and find a good correlation in all
Series only for the latter.  The correlation between $\langle P
\rangle_\rho$ and $R_{\rm mol}$ is potentially useful as a way to
estimate the typical internal pressure within gravitationally-bound
regions when only the total molecular-to-atomic mass is easily
accessible, as for low-resolution observations.  This internal
pressure is important in the small-scale aspects of star formation
such as determining the IMF \citep{2007ARA&A..45..565M}, as well as in
molecular chemistry.  The lack of correlation between $R_{\rm mol}$
and $\langle P \rangle_{\rm midplane}$ in Series K implies that the
molecular content cannot in general be predicted solely from $\Sigma$
and $\rho_{\ast}$ (i.e. without knowledge of $\kappa$), as noted above.
This reflects the fact that the formation of self-gravitating clouds
is regulated not just by gravitational processes and pressure, but
also by angular momentum.

5.  For our non-turbulent comparison models, we find that $R_{\rm
  mol}$ far exceeds observed values.  This indicates that turbulence
is essential to setting the observed phase balance in the ISM.
Recent theoretical investigations of the origin of Kennicutt-Schmidt
laws have focused on the dependence of star formation rates on the
molecular, rather than total, gas surface density
(e.g. \citealt
{2007arXiv0711.1361N,2008ApJ...680.1083R}).  Since turbulence is crucial in
determining the abundance of dense gas, in simulations that aim to
compute this abundance realistically it is necessary to incorporate
the feedback effects that drive turbulence, and to run on a fine
enough mesh (or with sufficient SPH particles) that the turbulence is well
resolved. While technically challenging in global disk models, local
models may offer a more immediate route to this goal.

{\it Caveats --} The models analyzed in this paper are subject to a
number of limitations, which could potentially affect some of our
conclusions. The chief limitations of the simulations are that (i)
they are two-dimensional, representing cuts in the $R-z$ plane, rather
than three-dimensional; (ii) we have adopted a very simple model to
implement turbulent driving as a star formation feedback effect from
HII regions, and we have not included other drivers of turbulence such
as supernovae, spiral shocks, and shear instabilities; (iii) we have
not included magnetic fields (or cosmic rays).  
We intend to pursue these extensions in future work.

Inclusion of magnetic fields and altered turbulent driving would
certainly affect the specific quantitative findings for $H_{\rm ave}$,
$\langle P \rangle_{\rm midplane}$, $\langle P \rangle_\rho$, and
$R_{\rm mol}$ in our models.  We believe, however, that the results we
have emphasized regarding physical {\it relationships} are robust.  In
particular, with appropriate modifications to include magnetic
stresses, the time-averaged vertical momentum flux through the
midplane must still equal
the time-averaged vertical weight if the mean vertical momentum is
conserved.  This can be used to predict the total
midplane pressure (including the magnetic pressure) and $H$ given the
values of $\Sigma$, $\rho_{\ast}$, and the thermal, turbulent, and
Alfv\'en velocities. Thus, we anticipate that inclusion of magnetic
fields and alternate turbulence sources would not fundamentally alter
the conclusion that reasonable estimates of scale heights can be
made using observable quantities even in highly-dynamic systems.

Further, we expect that our conclusions regarding the presence or
absence of correlations between $R_{\rm mol}$ and $\langle P
\rangle_\rho$ or $\langle P \rangle_{\rm midplane}$ would continue to
hold in models that include additional turbulence sources and magnetic
fields, although the details of correlations might change.  Namely,
angular momentum inherently must be important in permitting or
preventing formation of dense, self-gravitating clouds.  Our present
models account for angular momentum effects, and show that $R_{\rm
mol}$ does not in general have a one-to-one relationship with $\langle
P \rangle_{\rm midplane}$ or $\Sigma \sqrt{\rho_{\ast}}$; we expect this
result would carry over into any model that incorporates sheared
background rotation of the galactic disk.  Thus, if a one-to-one
relationship between $R_{\rm mol}$ and $\Sigma \sqrt{\rho_{\ast}}$ indeed
exists empirically, it implies that $\Sigma$, $\rho_{\ast}$, and $\kappa$
are not all independent quantities in real galaxies.

We are grateful to the referee for a number of comments that have helped 
improve our presentation.  Numerical computations used in this project
were carried out on the OIT High Performance Computing Cluster, and
the CTC cluster in the Department of Astronomy, at the University of
Maryland.  This work was supported by grant NNG05GG43G from NASA.

\bibliographystyle{apj}

\clearpage
\appendix
\section{Vertical Equilibrium with Stellar and Gas Gravity}

The vertical momentum equation, when averaged over a horizontal plane,
is given by 
\begin{equation}
\frac{\partial}{\partial t} \langle\rho v_z \rangle 
+ \frac{\partial   }{\partial z }
\left\langle P + \rho v_z^2 + \frac{
  {\bf B}\cdot{\bf B}}{8\pi} - \frac{ B_z^2 }{4\pi  } \right\rangle
= -\left\langle \rho \frac{\partial \Phi}{\partial z  }\right\rangle
\end{equation}
(see e.g. \citealt{2007ApJ...663..183P}).
Here, $\bf B$ is the magnetic field and $\Phi$ is the total (stellar
plus gaseous) gravitational potential.  In steady state 
$\langle\rho v_z \rangle$ is time-independent, so if we neglect
magnetic fields and assume that $\rho$, $v_z^2$, $c_s^2=P/\rho$, and 
$\partial \Phi/\partial z$ 
are statistically independent quantities, we obtain
\begin{equation}
\frac{ 1 }{\langle \rho\rangle}
\frac{\partial   }{\partial z }\left[\langle c_s^2 + v_z^2 \rangle \langle \rho \rangle\right]
= - \frac{\partial \langle \Phi\rangle}{\partial z }.
\end{equation}
The Poisson equation, also averaged over the horizontal plane and
assuming $R \Omega$ is independent of $R$, is 
\begin{equation}
\frac{\partial^2  \langle\Phi\rangle}{\partial z^2}
= 4 \pi G (\langle\rho\rangle + \rho_{\ast}),
\end{equation}
where $\rho_{\ast}$ is the background stellar density.

If we now define $\sigma_z^2 = \langle c_s^2 + v_z^2 \rangle$ and
assume that this total velocity dispersion is independent of height $z$,
we can combine the vertical momentum equation with the Poisson
equation to obtain  
a second-order differential equation for
the density profile $\langle \rho \rangle \rightarrow \rho(z)$:
\begin{eqnarray}
\frac{\partial}{\partial z}\left(
\frac{\sigma_z^2}{\rho(z)}\frac{\partial \rho(z)}{\partial z}
\right)
=-4\pi G\left(\rho_{\ast}+\rho(z)\right),
\label{eq:HSE}
\end{eqnarray}
Henceforth, we assume that $\rho_{\ast}$ is 
uniform within the gas disk, which is a good approximation provided
that the stellar scale height is significantly larger than the gaseous
scale height.  Equation (\ref{eq:HSE}) is the expression for effective
hydrostatic equilibrium in the vertical direction.

Introducing a variable $f(z)=\ln (\rho(z)/\rho_{\ast})$ and 
a constant $h^2=\sigma_z^2/(4\pi G\rho_{\ast})$,
we have
\begin{eqnarray}
f^{\prime\prime}=-\frac{1}{h^2}(1+e^f),
\end{eqnarray}
where the prime denotes a $z$ derivative. 
This can be integrated once as
\begin{eqnarray}
\frac{(f^{\prime})^2}{2}=-\frac{1}{h^2}(f+e^f)+\mbox{const}
=\frac{1}{h^2}\left(f_0-f+e^{f_0}-e^{f}\right),
\end{eqnarray}
where $f_0=\ln(\rho_0/\rho_{\ast})$ 
is the boundary condition at the midplane where $f^{\prime}=0$.
If we Taylor expand and retain only the two lowest order terms, i.e.
$f(z)=f_0 - f_1 z^2$ so that 
$\rho/\rho_0= \exp(-f_1 z^2)$, 
the governing ODE becomes an algebraic equation: 
\begin{eqnarray}
\frac{(2f_1 z)^2}{2}
=\frac{z^2}{h^2}\left(f_1+\frac{\rho_0}{\rho_{\ast}}f_1\right) 
=\frac{4\pi G(\rho_0+\rho_{\ast})}{\sigma_z^2}f_1 z^2.
\end{eqnarray}
The coefficient $f_1$ is 
\begin{eqnarray}
f_1=\frac{1}{2H^2},\quad 
H^2=\frac{\sigma_z^2}{4\pi G(\rho_{\ast}+\rho_0)}.
\label{eq:Height}
\end{eqnarray}
Therefore, the gas density and pressure are 
approximately given by Gaussian profiles
\begin{eqnarray}
\rho(z)=\rho_0 e^{\displaystyle -\frac{z^2}{2H^2}},\quad
 P(z)=P_0e^{\displaystyle -\frac{z^2}{2H^2}},
\end{eqnarray}
where $P_0=\sigma_z^2\rho_0$.
The midplane gas density $\rho_0$ is determined by requiring that 
the profile integrates to the (known) gas surface density,
\begin{eqnarray}
\Sigma=\int_{-\infty}^{\infty}\rho(z)\, dz=\sqrt{2\pi}\rho_0 H.
\end{eqnarray}
Substituting for $\rho_0$ 
in equation (\ref{eq:Height}), the scale height $H$ must satisfy 
\begin{eqnarray}
H^2&=&\frac{\sigma_z^2}{4\pi
 G(\rho_{\ast}+\frac{\Sigma}{\sqrt{2\pi}H})}.
\end{eqnarray}
This yields a quadratic equation for $H$, with solution given
by equation (\ref{Hest_eq}) of the text.

\end{document}